\documentclass[journal,onecolumn]{IEEEtran}
\usepackage{cite}
\usepackage{graphicx}
\usepackage[cmex10]{amsmath}
\usepackage{algorithm}
\usepackage{algorithmic}
\usepackage{booktabs}
\usepackage{array,amssymb,multirow,dcolumn}
\usepackage{epstopdf}
\usepackage{color}
\usepackage[tight,footnotesize]{subfigure}
\usepackage[table]{xcolor}
\usepackage{textcomp}
\usepackage{url}
\usepackage{subfigure}

\usepackage{color}
\begin{document}
	
	\title{PMU-Based ROCOF Measurements:\\Uncertainty Limits and Metrological Significance\\in  Power System Applications}
	
	\author{G. Frigo, A. Dervi{\v s}kadi{\'c}, Y. Zuo, and M. Paolone
		\thanks{Manuscript received X Month 2018.}
		\thanks{Corresponding author's e-mail: guglielmo.frigo@epfl.ch.}
		\thanks{G. Frigo, A. Dervi{\v s}kadi{\'c}, Y. Zuo, and M. Paolone are with \'Ecole Polytechnique F\'ed\'erale de Lausanne, 1015 Lausanne, Switzerland.}
		\thanks{Accepted for publication on IEEE Transactions on Instrumentation and Measurement on 9 March 2019.}}
	\maketitle
	
	\begin{abstract}
		In modern power systems, the Rate-of-Change-of-Frequency (ROCOF) may be largely employed in Wide Area Monitoring, Protection and Control (WAMPAC) applications. However, a standard approach towards ROCOF measurements is still missing. In this paper, we investigate the feasibility of Phasor Measurement Units (PMUs) deployment in ROCOF-based applications, with a specific focus on Under-Frequency Load-Shedding (UFLS). For this analysis, we select three state-of-the-art window-based synchrophasor estimation algorithms and compare different signal models, ROCOF estimation techniques and window lengths in datasets inspired by real-world acquisitions. In this sense, we are able to carry out a sensitivity analysis of the behavior of a PMU-based UFLS control scheme. Based on the proposed results, PMUs prove to be accurate ROCOF meters, as long as the harmonic and inter-harmonic distortion within the measurement pass-bandwidth is scarce. In the presence of transient events, the synchrophasor model looses its appropriateness as the signal energy spreads over the entire spectrum and cannot be approximated as a sequence of narrow-band components. Finally, we validate the actual feasibility of PMU-based UFLS in a real-time simulated scenario where we compare two different ROCOF estimation techniques with a frequency-based control scheme and we show their impact on the successful grid restoration.
	\end{abstract}
	

	\section{Introduction}
	In modern power systems, the frequency first time-derivative associated to the signal fundamental component, also referred to as Rate-of-Change-of-Frequency (ROCOF), may be largely employed in Wide Area Monitoring, Protection and Control (WAMPAC) applications, like load shedding \cite{Amraee-etAl2017}, islanding detection \cite{Gupta-etAl2017}, and distributed generation control \cite{Freitas-etAl2005}. However, a standard approach towards ROCOF measurement is still missing: most applications adopt estimation algorithms and signal models designed \textit{ad hoc} to suitably meet the operating conditions and requirements \cite{Riepnieks-etAk2016}.
	
	The recent literature has been considering the adoption of Phasor Measurement Units (PMUs) in ROCOF-based applications because of two main advantages.
	First, PMUs are able to perform ROCOF measurements characterized by fast reporting rates and responsiveness \cite{Ding-etAl2016}, as well as strict accuracy limits. In this regard, in compliance with the IEEE Std. C.37.118.1 (IEEE Std), PMUs are expected to provide an updated ROCOF estimate at reporting rates of tens of frames per second (fps), limiting the uncertainty level to 0.01 Hz/s in steady-state conditions, 6 Hz/s in the presence of harmonic distortion, and 3 Hz/s in dynamic conditions \cite{ieeeC37}. Second, PMUs allow for a distributed measurement infrastructure that provides a grid-state awareness by means of synchronous monitoring of voltage and current phasors in different grid nodes \cite{Kamwa-etAl2011,Pignati-etAl2015,vonMeier-etAl2017}. 
	
	Based on the adopted synchrophasor estimation technique, it is possible to identify four main PMU algorithmic classes: demodulation, window-based, time-domain, and recursive filtering \cite{Paolone-etAl2017}. In particular, for the sake of applicability and of replicability of results, we focus on window-based algorithms, that divide the signal into partially overlapped finite-length segments and consider their spectral representation by computing the Discrete Fourier Transform (DFT).
	In this context, many factors limit the actual accuracy of ROCOF estimates, as provided by PMUs. First, the IEEE Std models the signal fundamental component through a synchrophasor representation in the frequency domain: the signal DFT can be approximated by a term associated to the fundamental component, plus a restricted set of coefficients associated to eventual harmonic or inter-harmonic contributions. However, this finite-spectrum model cannot account for rapid variations of signal parameters with a satisfying level of accuracy \cite{Phadke-etAl2009}. In this case, a parametric ROCOF estimation (as provided by window-based techniques) becomes model-dependent and cannot be considered as a unique property of the signal \cite{Roscoe-etAl2017}.
	
	Second, current synchrophasor- and ROCOF-based applications should comply with different requirements in terms of update rate and measurement reliability. Typically, PMUs consider reduced window lengths to keep the reporting latency within some tens of milliseconds, whereas ROCOF-based relays adopt longer windows \cite{Gupta-etAl2017}. 
	It is worth pointing out that PMUs instead do not employ averaging to limit the uncertainty contributions due to measurement noise \cite{Macii-etAl2016}, unbalance \cite{Castello-etAl2018}, and signal distortion \cite{Dickerson-2015}.
	Based on these considerations, the recent IEEE Std amendment has significantly relaxed ROCOF error (RFE) requirements: the harmonic distortion test has been suspended for class M, whereas the class P limit has been relaxed up to 0.4 Hz/s \cite{ieee_am}. In practice, though, such uncertainty level makes ROCOF estimates totally unreliable and, thus, useless for the aforementioned applications \cite{Derviskadic-etAl2018}. In this regard, it is interesting to consider that real-world WAMPAC applications might deal with sudden variations and ROCOF values largely exceeding the IEEE Std limits \cite{aemo_report}.
	
	In a previous publication \cite{Frigo-etAl2018}, we compared the ROCOF estimation accuracy as provided by three window-based algorithms \cite{Bertocco-etAl2015, Romano-etAl2014, Derviskadic-etAl2017} taken from the recent synchrophasor estimation literature. In this context, we evaluated how static and dynamic signal models represent plausible time-varying ROCOF trends, inspired from real experimental acquisitions.
	This paper further evaluates how different ROCOF computation techniques can influence the actual uncertainty of ROCOF measurements. Indeed, since PMU-based measurements prove to be signal model-dependent, it is necessary to carry out a sensitivity analysis, which takes into account non only the algorithm parameters, but the cumulative non-linear effect of synchrophasor estimation, as function of the network dynamic response and adopted control scheme. As a consequence, for this analysis, we do not reproduce the static and dynamic test conditions defined in IEEE Std, but realistic operating conditions, inspired by real-world 
	networks, like the frequency excursions typical of islanded grids (e.g. Hydro-Quebec grid \cite{Trudel-etAl2005}), the effect of large inter-area oscillations in the Pan-Europe transmission network \cite{SiL-report}, and the islanding maneuver of an active distribution network characterized by reduced inertia \cite{Borghetti-etAl2011}. It is worth to point that the two last datasets were already considered in \cite{Frigo-etAl2018}, whereas the first one was specifically synthesized for the present publication.
	
	As an extension of the above-mentioned research, this paper further investigates the feasibility of PMU-based ROCOF measurement in a real-world operating scenario, i.e. an Under-Frequency Load-Shedding (UFLS) application \cite{Derviskadic-etAl2018}. In reduced-inertia power systems, UFLS techniques allow for minimizing the risk of uncontrolled separation, loss of generation, or shutdown, when the energy demand exceeds what the primary power source can supply \cite{ieee-gui}. Traditionally, the amount and the sequence of loads to be shed is determined based only on frequency measurements \cite{Cote-etAl2001,Tang-etAl2013,Laverty-etAl2015}. Nevertheless, recent literature has suggested the adoption of ROCOF-based relays to guarantee a prompter and more effective response in the presence of fast variations \cite{Arulampalam-etAl2010,Mokari-etAl2014}.
	In that sense, the simulation platform presented in \cite{Derviskadic-etAl2018}, representing a ROCOF-based UFLS plan, can be regarded as a benchmark for testing the effectiveness of ROCOF estimations in power systems applications.
	
	In this regard, we also present a real-time simulation of an UFLS scheme, where we employ both static and dynamic model-based PMUs, and we compare them with a traditional frequency-based control scheme. The obtained results confirm how ROCOF-based relays might represent an effective solution for a successful grid restoration.
	
	The paper is organized as follows. In Section II, we introduce the processing approach of current ROCOF-based relays and the comparison with PMU-based ROCOF measurements. Section III briefly describes the three synchrophasor estimation algorithms and presents the considered ROCOF estimation techniques. In Section IV, we introduce the parameters of the numerically simulated scenarios: we characterize the algorithms' performance (in terms of RFE) in three different datasets, inspired by real-world experimental acquisitions. Section V presents a real-time simulated scenario of an UFLS application relying on PMU-based ROCOF measurements. Finally, in Section VI we provide some closing remarks.
	
	\section{ROCOF-based Relay}
	In this Section, we briefly describe the traditional approach towards ROCOF measurement, as implemented by most modern ROCOF-based relays \cite{ieee-gui}.
		
	\begin{figure}
		\centering
		\includegraphics[width=.5\columnwidth]{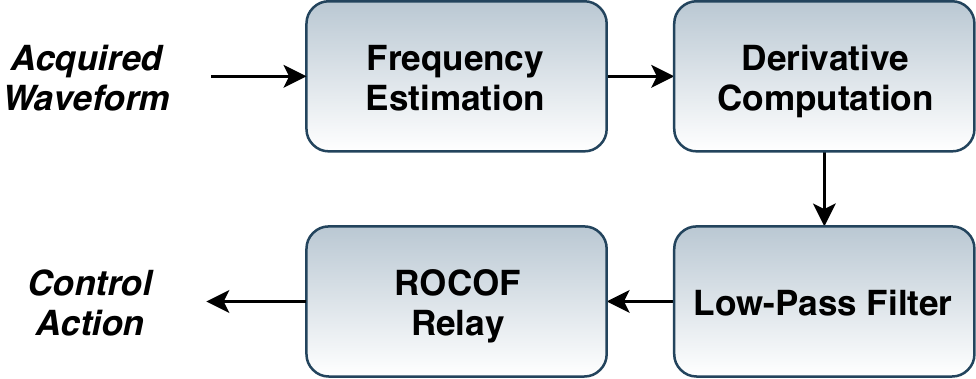}
		\caption{Block scheme of the typical operational sequence of a ROCOF relay employed in power system applications \cite{dnv_report}.}
		\label{fig:rocof_relay}
	\end{figure}

	In the context of transmission networks, Fig. \ref{fig:rocof_relay} represents the main processing stages of a generic control scheme based on the ROCOF estimates \cite{dnv_report}:
	\begin{itemize}
		\item first, we extract the fundamental frequency information. Typically, this stage employs an extended window length, in the order of seconds, in order to optimize the frequency resolution and estimation accuracy;
		\item then, we compute ROCOF as the first-order time-derivative of frequency, typically implemented as the incremental ratio between consecutive frequency estimates;  
		\item a low-pass filtering stage allows for removing fast ROCOF dynamics and thus providing a smoother trend. In general, we apply a moving average filter, whose window length has to be suitably scaled based on the expected variation range and bandwidth of ROCOF estimates;
		\item finally, we compare the obtained measurements with given threshold levels (based on the specific network inertia properties), whose excess produces the control action to be activated. 
	\end{itemize}
	
	Typically, ROCOF relay performance is characterized in terms of measurement resolution and detection time \cite{Vieira-etAl2006}. In this sense, the adoption of extended window lengths and averaging filters might improve the ROCOF estimation, but significantly deteriorate the responsiveness.  
	In the time-sensitive scenario of ROCOF-based applications \cite{Wall-etAl2016}, PMUs might represent a promising measurement infrastructure to provide synchronized estimates, with high estimation accuracy and low reporting latency.
	
	Indeed, in a PMU-based measurement scenario, the IEEE Std requirements affect or even prevent the implementation of the same processing procedure. First of all, the limitations in terms of reporting latency make difficult to perform the frequency (and ROCOF) estimation over window length that exceed three and five nominal cycles, i.e. 60 and 100 ms at 50 Hz, for P- and M-class, respectively \cite{ieeeC37}.
	
	As regards the estimation accuracy, the IEEE Std requires RFE to not exceed 0.01 Hz/s in steady-state conditions, and 6 Hz/s in the presence of harmonic distortion \cite{ieeeC37,ieee_am}. However, in order to apply ROCOF measurements in real-world scenarios, we need a metering infrastructure more resilient against interfering components and characterized by a wider dynamic range. In other words, the PMU should be able to provide reduced RFE independently from the variation speed of the fundamental frequency or the distortion level \cite{Kamwa-etAl2014}.
	
	Furthermore, the PMU relies on the synchrophasor signal model, that assumes the fundamental component to be approximated by a narrow-band spectral component. During transient events, though, the power signal is characterized by a continuous spectrum, and an analysis limited to the PMU pass-bandwidth provides only a partial and inconsistent representation of the grid state\footnote{If the signal under investigation cannot be even approximated as periodic and band-limited, even the Fourier series representation looses its significance.}. For this reason, it is reasonable to say that PMU-based measurements are not reliable in the correspondence of sudden amplitude or phase changes, like islanding or fault events \cite{Barchi-etAl2013b}.
	
	It should also be noticed that PMUs are metering devices specifically designed to produce frequently updated measurements with a reporting rate in the order of tens fps. Typically, they consider short window lengths and they do not apply any moving average compensation. As a consequence, PMU estimates account for the quasi-instantaneous voltage and current variations, but their accuracy tends to deteriorate in the presence of dynamic trends or disturbances.
	
	\section{Selected ROCOF Estimation Algorithms}
	The IEEE Std refers to synchrophasor measurements in power systems, and introduces the accuracy and latency requirements for PMUs in terms of synchrophasor, frequency and ROCOF estimation. In this regard, ROCOF is defined as the estimated frequency first derivative associated to the signal fundamental component. In the IEEE Std Annex C, two signal processing models are presented as the reference estimation methods for class M and P, respectively. Both classes compute ROCOF as the incremental ratio between two consecutive frequency estimates \cite{ieeeC37}.
	
	In this paper, we compare the accuracy and reliability provided by four different ROCOF estimation techniques, based on three consolidated state-of-the-art algorithms, namely the Compressive Sensing-based Taylor-Fourier Model (cs-TFM) \cite{Bertocco-etAl2015}, the Enhanced Interpolated DFT (e-IpDFT) \cite{Romano-etAl2014}, and the Iterative Interpolated DFT (i-IpDFT) \cite{Derviskadic-etAl2017}. The proposed analysis enables us to evaluate the ROCOF estimates' dependence from the algorithm parameters, like the adopted signal model and window length.
	
	It is worth noticing that the selected algorithms are representative of window-based approaches relying on static and dynamic synchrophasor formulations. On one side, cs-TFM relies on a dynamic signal model that accounts also for the frequency first time-derivative, and is thus capable to estimate ROCOF along with the other fundamental component parameters, i.e. amplitude, initial phase and frequency. 
	On the other side, both IpDFT approaches rely on a static signal model, and exploit interpolation techniques to enhance the frequency resolution and reduce spectral leakage effects. In particular, e-IpDFT algorithm applies an iterative routine to mitigate the long-range spectral leakage due to the image of the fundamental component at the negative frequency, whereas i-IpDFT attempts to compensate also the short-range spectral leakage caused by harmonic or inter-harmonic components.
	
	Static estimation techniques, like e-IpDFT and i-IpDFT, can compute ROCOF only as the incremental ratio between two consecutive frequency estimates. 
	It is thus reasonable to expect that the ROCOF estimates are partially delayed and smoothed depending on the adopted reporting period. 
	Conversely, the dynamic estimation techniques, like cs-TFM, enable us to directly compute the instantaneous ROCOF, but it is still possible to apply also the finite difference formulation, in coherence with the static estimation techniques. The adoption of a dynamic signal model allows for tracking possible time-varying trends, but at the same time suffers from higher sensitivity to uncompensated disturbances \cite{Castello-etAl2012,Narduzzi-etAl2018}. 
	 
	\section{ROCOF Error Analysis}
	
	In this Section, we characterize the estimation accuracy of the considered algorithms in three datasets, that reproduce realistic operating scenarios. For this analysis, we consider waveforms inspired by real-world experimental acquisitions, rather than IEEE Std test conditions, as available in \cite{Bertocco-etAl2015, Romano-etAl2014, Derviskadic-etAl2017}.
	
	In theory, it is possible to interpret any synchrophasor estimation algorithm as a generic unbiased estimator. In this sense, it is possible to define the expected accuracy limits in the presence of uncertainty contributions, like measurement noise or phase unbalance. As regards the measurement noise, the worst-case RFE depends on the Equivalent Noise Bandwidth (ENBW) of the adopted window \cite{Macii-etAl2016}. Similarly, as regards the phase unbalance, the ROCOF estimation accuracy depends on the algorithm capability of identifying and mitigating the equivalent harmonic disturbances \cite{Castello-etAl2018}. By considering these uncertainty sources as separate and independent contributions, the dynamic estimation techniques are proven to outperform the static counterparts. In this paper, instead, we carry out a performance assessment in real-world operating conditions, where several non-linear effects are mixed together, like distortion and inter-modulation effects, network dynamic behavior and control schemes.
	
	For this analysis, we numerically simulate in Matlab a plausible synchrophasor estimation context, with sampling frequency and reporting rate equal to 5 kHz and 50 fps, respectively. Given a nominal system frequency of 50 Hz, we consider two window lengths, namely 60 and 100 ms, as representative of protection and measurement PMU application classes. In particular, the three-cycle window proves to be compliant with class P requirements in terms of reporting latency\footnote{PMU estimates are typically referred to the observation interval midpoint. Therefore, in order to meet the P-class latency requirement of 40 ms, the window length cannot exceed four nominal cycles, i.e. 80 ms.}, whereas longer window lengths are typically associated to M-class applications.
	
	The M-class configuration is expected to outperform the P-class counterpart in terms of estimation accuracy, as a longer window length corresponds to an enhanced frequency resolution. Nevertheless, we include also P-class configuration in this analysis in order to assess whether UFLS application might benefit from their faster response in relay's operation.
	
	For each dataset, we characterize the RFEs provided by both static (e-IpDFT, i-IpDFT) and dynamic (cs-TFM) approaches. In this last case, we compare the results obtained with derivative and finite-difference formulations, in order to assess which one provides the most reliable ROCOF estimation.
	
	The proposed analysis provides a statistical description of the considered algorithms' estimation accuracy. For this reason, we compute the RFE cumulative distribution function (CDF) and we characterize it in terms of mean value, standard deviation and 95th quantile (as the errors are not normally distributed). In this way, we are able to determine not only the estimator worst-case performance, but also its range of variability. In the presence of small ROCOF variations, though, a reduced RFE is not sufficient to guarantee a sufficient resolution capability. As a consequence, we compute also the Pearson's correlation coefficient between the estimated and ground-truth values. In this way, we assess whether the estimation technique operates as a linear predictor and the ROCOF estimates preserve the original information content.
	
	\paragraph{Dataset I} The first dataset has been inspired by a real-world operating scenario: the Hydro-Québec grid \cite{Cote-etAl2001}. In fact, this grid represents a significant test-bench for evaluating the robustness and accuracy of the synchrophasor estimation algorithms and ROCOF-based applications as it presents many technical and processing challenges \cite{Kamwa-etAl2014}. In particular, the Hydro-Québec grid is characterized by large frequency excursions and affected by several harmonic and inter-harmonic interferences (further details in \cite{Trudel-etAl2005}).
	
\begin{figure}
		\centering
		\includegraphics[width=.5\columnwidth]{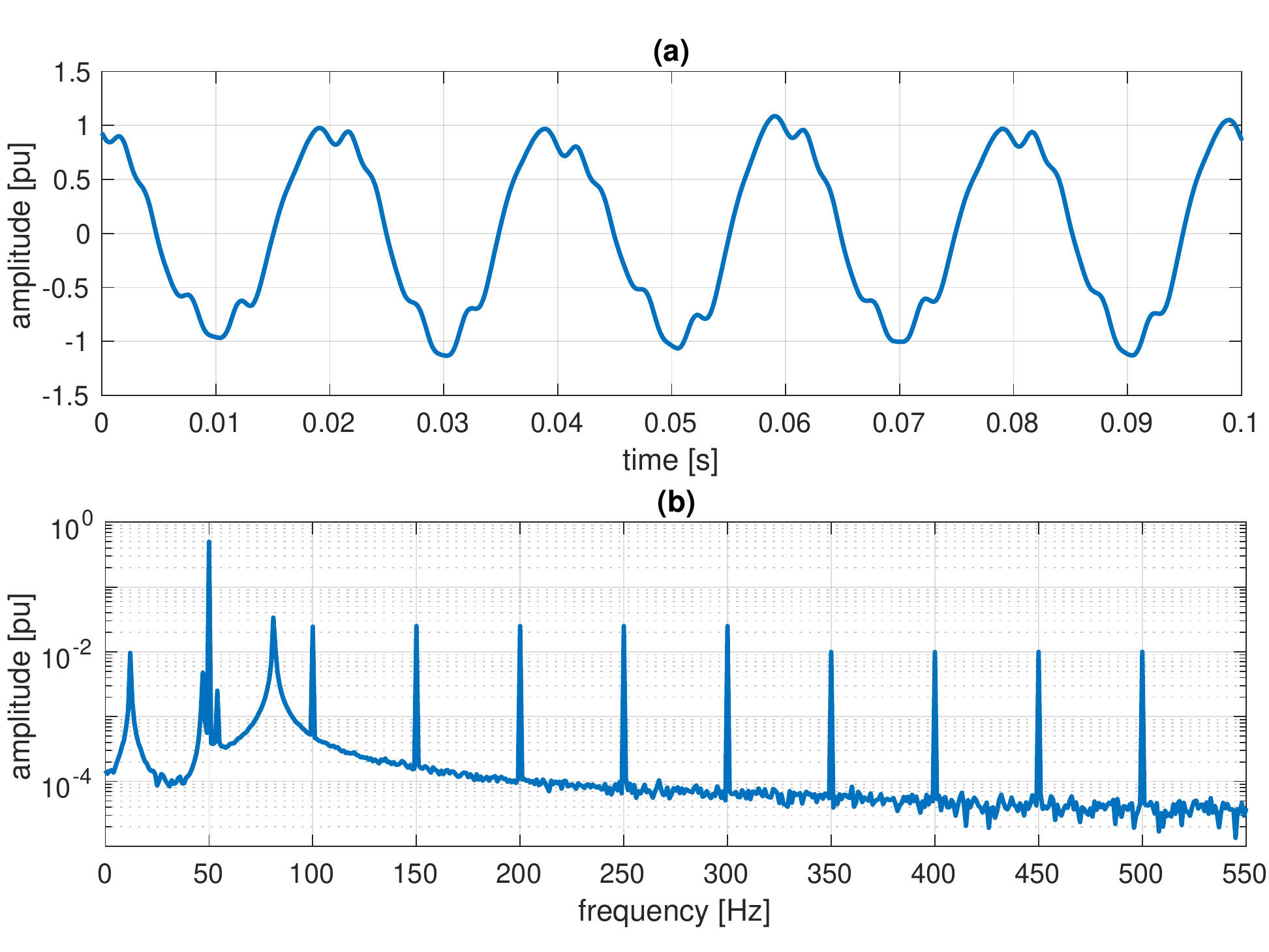}
		\caption{Representation of the simulated test waveform in time- and frequency-domain in (a) and (b), respectively. The measurement noise is reproduced by an uncorrelated white Gaussian variable, such that SNR  is equal to 60 dB.}
		\label{fig:hydro_time}
	\end{figure}

	Based on the signal model proposed in \cite{Kamwa-etAl2013}, we design a test waveform that presents a realistic variation trend of the fundamental frequency and ROCOF\footnote{For the sake of consistency with the other datasets, we consider a system frequency of 50 Hz, and we accordingly scale the inter-harmonic and inter-modulation component frequencies. Comparable results can be obtained in the original Hydro-Québec scenario, where the system frequency is 60 Hz.}. As shown in Fig. \ref{fig:hydro_time}(a), in the time-domain the waveform is characterized by a significant distortion level with THD and SINAD equal to 11.18\% and 16.94 dB, respectively, as result of the interference components visible in the spectral representation of Fig. \ref{fig:hydro_time}(b).
	
	In this case, we define the test waveform as the combination of a steady-state fundamental component with different kinds of disturbances, whose harmonic indexes and normalized amplitudes are reported in Table \ref{tab:hydro_wav_comp}.
	
	\begin{table}
		\centering
		\caption{Dataset I: Waveform Components Parameters}
		\label{tab:hydro_wav_comp}
		\begin{tabular}{c|cc}
				\toprule
				\textbf{Component} & \textbf{Harm. Index} & \textbf{Norm. Amplitude}\\
				\midrule
				fundamental & 1 & 1 \\
				& &\\
				\multirow{2}{*}{inter-modulation} & 0.936 & 0.01\\
				& 1.082 & 0.005\\
				& &\\
				\multirow{2}{*}{harmonics} & from 2 to 6& 0.05 \\
				& from 7 to 10 & 0.02 \\
				& &\\
				inter-harmonic & 1.625 & 0.075\\
				& &\\
				sub-harmonic & 0.243 & 0.02\\
				\bottomrule
		\end{tabular}
	\end{table}

\begin{figure}
		\centering
		\includegraphics[width=.5\columnwidth]{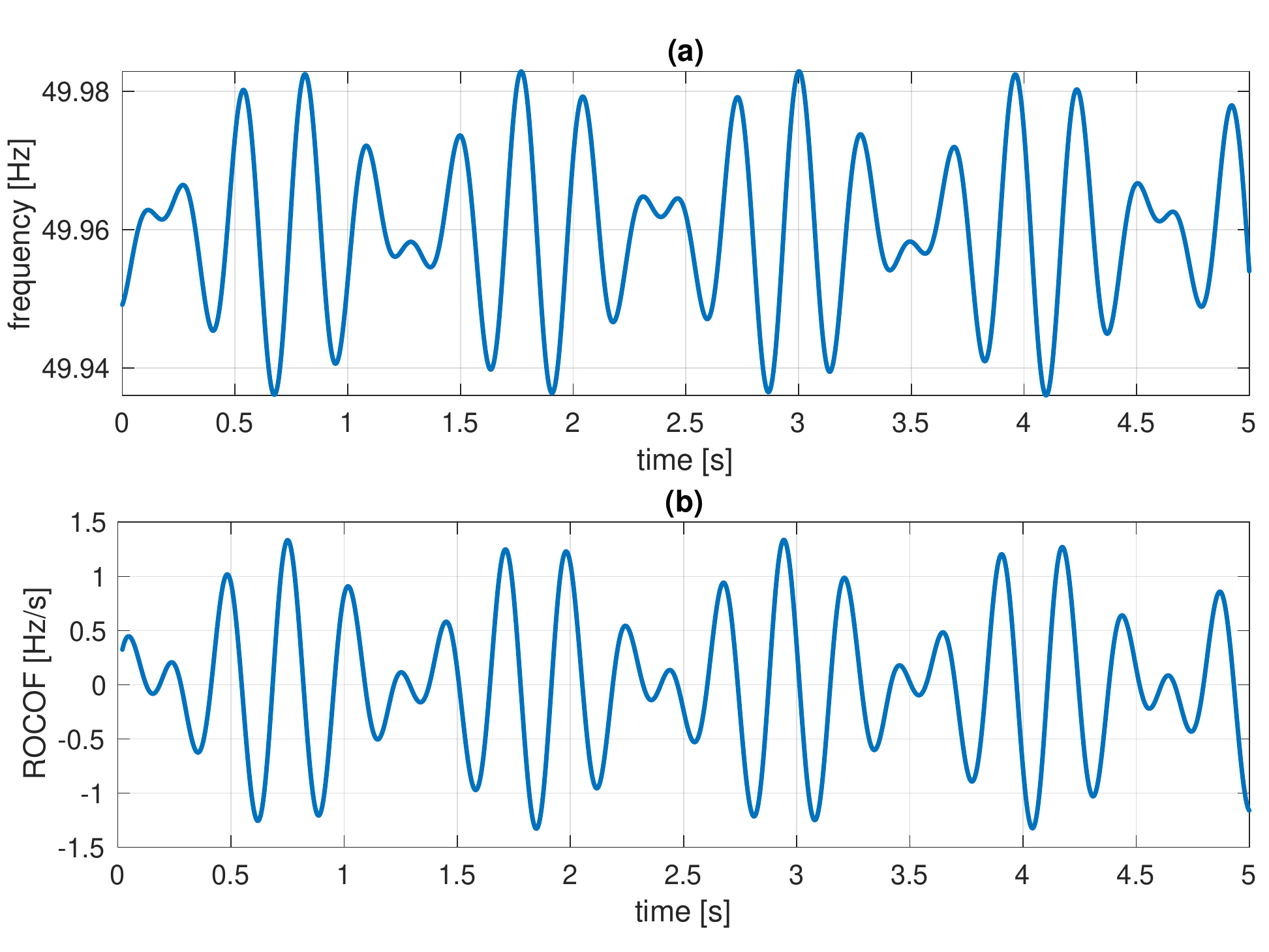}
		\caption{In (a), the ground-truth frequency associated to the signal fundamental component. In (b), the corresponding ROCOF defined as finite-difference derivative of two consecutive estimates at intervals of 20 ms.}
		\label{fig:hydro_rocof}
	\end{figure}

	It is worth noticing that the fundamental tone is surrounded by two asymmetrical inter-modulation components, that differ in terms of normalized amplitude and deviation with respect to the system frequency. Since these three components lay within the fundamental range [45, 55] Hz, it is reasonable to expect that the PMU cannot distinguish them as separate tones, but only measure their combined contribution \cite{Castello-etAl2016}. In this regard, Fig. \ref{fig:hydro_rocof}(a) presents the instantaneous frequency associated to the sum of fundamental and inter-modulation components. In fact, given the cosine sum formula, we are able to define the instantaneous frequency at each sampling time, i.e. at each 200 $ \mu $s. Coherently with the adopted reporting rate of 50 fps, we define the corresponding ground-truth of ROCOF value as the incremental ratio between two consecutive frequency estimates at intervals of 20 ms.
	
	In addition, the fundamental tone is also affected by narrow-band disturbances, namely nine harmonics (up to 500 Hz), an inter-harmonic at 81.25 Hz, and a sub-harmonic at 12.15 Hz. In this case, the spectral components are located outside the PMU pass-bandwidth [25, 75] Hz, and thus the PMU is expected to reject their injections in its final ROCOF estimates.
	
	Finally, in order to model the measurement uncertainty, we corrupt the test waveform with an additive and uncorrelated white Gaussian noise, characterized by a SNR of 60 dB, that corresponds to a resolution of nearly 10 bits.
	
	In summary, the simulated scenario presents two main challenges in terms of ROCOF measurements. First, the ground-truth ROCOF exhibits fast and irregular oscillations within a rather wide variation range, i.e. $ \pm $1.5 Hz/s. Second, the estimation of fundamental component parameters is affected by the spectral leakage caused by harmonic, sub- and inter-harmonic components. For this reason, the adopted signal model can be considered as a representative test-bench for ROCOF estimation techniques, even if it neglects some features of the original Hydro-Québec grid.
	
	For this analysis, we consider a waveform duration of 5 s. Given a reporting rate of 50 fps, the P- and M-class configurations produce 248 and 246 ROCOF estimates, respectively. By comparing the estimated values with the previously defined ROCOF ground-truth, we are able to determine the estimation error RFE and infer its statistical properties. In this regard, since we do not know a priori the RFE statistical distribution, we compare the error CDFs as provided by each estimation technique.
	
\begin{figure}
		\centering
		\includegraphics[width=.5\columnwidth]{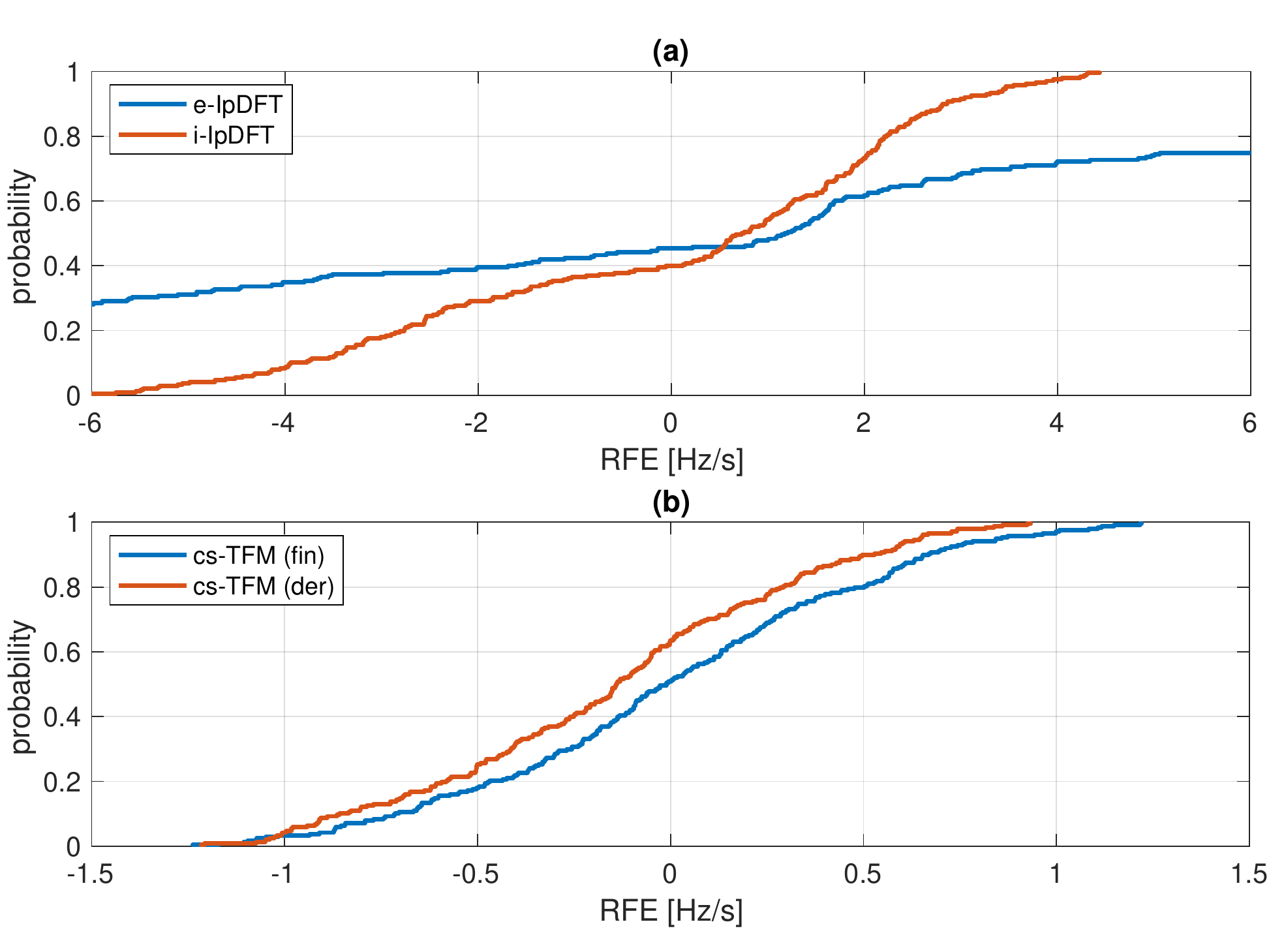}
		\caption{Given a window length of 60 ms (P-class configuration), comparison of the cumulative distribution functions associated to the RFEs as produced by static (a) and dynamic model-based algorithms (b).}
		\label{fig:hydro_cdf_3c}
	\end{figure}

	In Fig. \ref{fig:hydro_cdf_3c} we consider the error CDFs associated to the P-class configuration for static (a) and dynamic (b) approaches. In the static case, both algorithms do not succeed in rejecting the spectral leakage effects from interfering tones. Indeed, the harmonic and interharmonic content is so severe that the DFT bins related to the fundamental tone are largely biased \cite{Derviskadic-etAl2017}.
	As a result, the maximum RFE is limited to 11 and 6 Hz/s for e-IpDFT and i-IpDFT, respectively, i.e. it exceeds the expected variation range by almost one order of magnitude.
	In the dynamic case, instead, the performance is rather independent from the adopted ROCOF formulation (either finite difference or derivative). However, the obtained RFEs still exceed 1 Hz/s.
	
	Besides providing the statistical representation characteristics by means of CDF, we also synthetically present the error distribution parameters in Table \ref{tab:hydro_result}, in terms of error mean, standard deviation, and 95-th percentile. 
	In addition, we compute also the Pearson's correlation index between estimated and ground-truth ROCOF, that is limited to just 2.64\% for P-class best case. As a consequence, even if cs-TFM approaches guarantee a performance enhancement in terms of maximum error, the corresponding estimates do not provide any information regarding the ground-truth ROCOF.
	
	\begin{table}
		\centering
		\caption{Dataset I: RFE Statistical Properties}
		\label{tab:hydro_result}
		\begin{tabular}{c|c||ccc|c}
				\toprule
				\textbf{Estimator} & \textbf{Class} & \textbf{Mean} & \textbf{Std Dev} & \textbf{95\%} & \textbf{Corr}\\
				\midrule
				\multirow{2}{*}{e-IpDFT} & P & -2.45$ \cdot $10$ ^{-2} $ & 6.76 & 10.51 & 0.77\% \\
				& M & 6.25$ \cdot $10$ ^{-4} $ & 0.89 & 2.03 & 50.31\%\\
				\midrule
				\multirow{2}{*}{i-IpDFT} & P & -1.90$ \cdot $10$ ^{-2} $ & 2.63 & 5.59 & 2.22\%\\
				& M & 1.73$ \cdot $10$ ^{-4} $ & 0.28 & 0.57 & 88.78\% \\
				\midrule
				cs-TFM & P & -2.10$ \cdot $10$ ^{-3} $ & 0.52 & 1.12 & 2.60\%\\
				(fin) & M & 2.92$ \cdot $10$ ^{-4} $ & 0.15 & 0.38 & 96.29\% \\
				\midrule
				cs-TFM & P & -1.51$ \cdot $10$ ^{-1} $ & 0.48 & 1.11 & 2.64\%\\
				(der) & M & 1.20$ \cdot $10$ ^{-2} $ & 0.21 & 0.56 & 92.59\% \\
				\bottomrule
		\end{tabular}
	\end{table}
	
\begin{figure}
		\centering
		\includegraphics[width=.5\columnwidth]{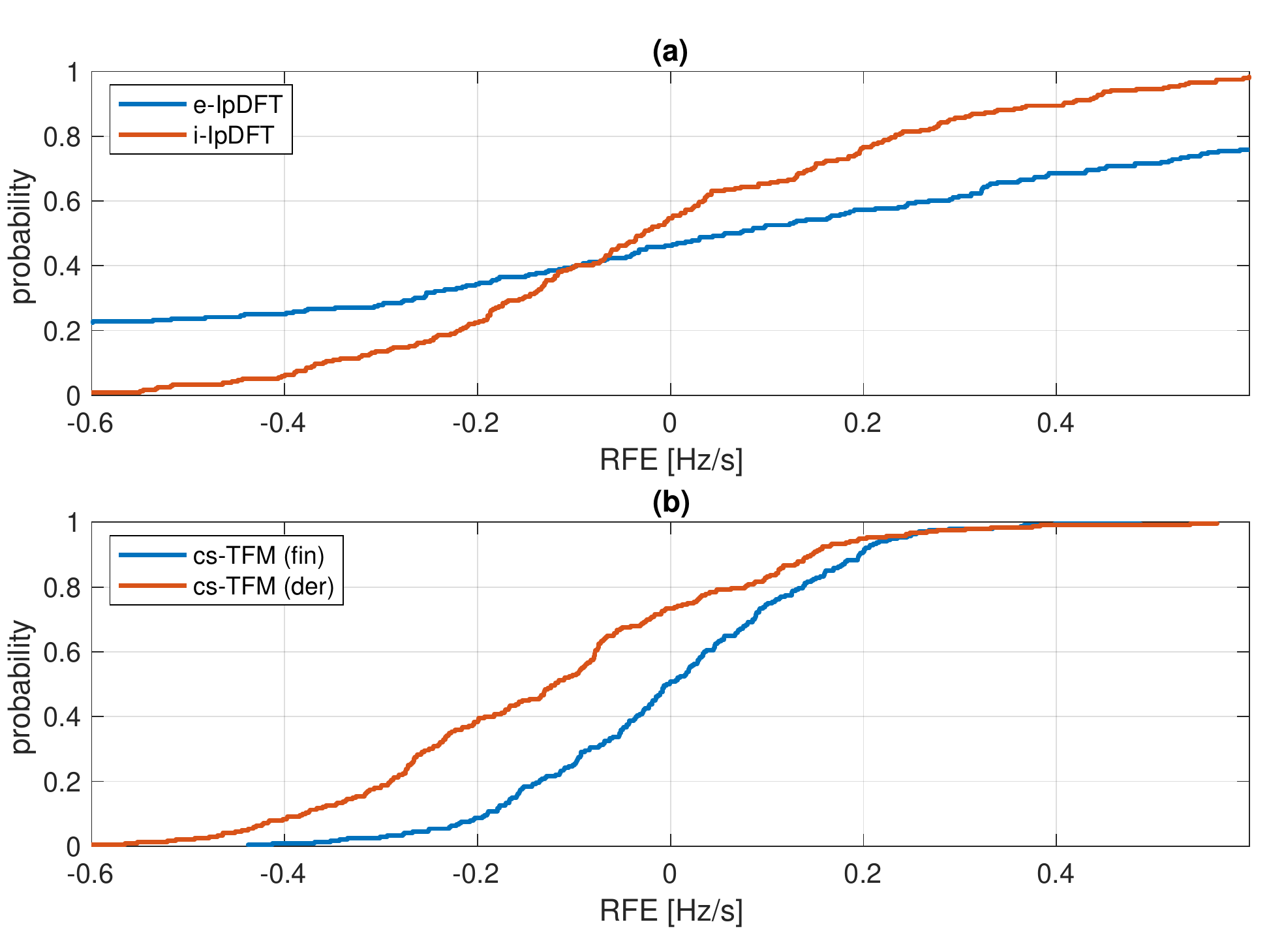}
		\caption{Given a window length of 100 ms (M-class configuration), comparison of the cumulative distribution functions associated to the RFEs as produced by static (a) and dynamic model-based algorithms (b).}
		\label{fig:hydro_cdf_5c}
	\end{figure}

	Fig. \ref{fig:hydro_cdf_5c} represents the error CDFs associated to the M-class configuration for static (a) and dynamic (b) approaches. The increased window length corresponds to an enhanced frequency resolution and a reduced spectral leakage. Accordingly, it is reasonable to expect a significant improvement for all the considered estimation techniques.
	
	In the upper graph, the i-IpDFT mitigates the injections from sub- and inter-harmonic disturbances, and thus reduces the error range up to $ \pm $0.7 Hz/s. Conversely, e-IpDFT compensates only the leakage contribution of the fundamental image at negative frequency. As a results, the corresponding maximum estimation error still exceeds 2.1 Hz/s.
	
	In the lower graph, instead, the cs-TFM dynamic model allows for a more accurate tracking of the ROCOF time-varying trend, with an error range limited to $ \pm $0.5 Hz/s. It is also worth noticing that the finite difference formulation provides a further performance enhancement since its inherent smoothing effect limits the spurious oscillations induced by the interfering components, whereas the derivative formulation is more easily affected by uncompensated disturbances.
	
	Similar considerations hold also for the correlation index between estimated and ground-truth values. Among the static approaches, the i-IpDFT outperforms the e-IpDFT and achieves a correlation of 88.78\% thanks to the mitigation of spectral leakage due to out-of-band interferences. On the other hand, the dynamic approaches provide a nearly optimal result, with a correlation exceeding 96\% for both formulations. Accordingly, it is reasonable to say that the dynamic signal model allows for better tracking the ROCOF time-varying trend. In more detail, the best performance is provided by the finite difference formulation, as it partially filters out the residual spurious injections from the interfering components.
	
	\paragraph{Dataset II} The second dataset refers to a real-world network event, recorded on December 1, 2016, when an unexpected opening of a line in the French transmission network caused an inter-area oscillation in the Continental Europe electricity system \cite{SiL-report}. In particular, the test waveform derives from the estimates of a PMU employed by the grid operator of the city of Lausanne, i.e. \textit{Services industriels de Lausanne} (SiL). Further details are available in \cite{Derviskadic-etAl2016}.
	
	Based on PMU estimates of fundamental frequency, amplitude and initial phase, we can recover the time-domain waveform as sampled at 5 kHz, through the approach described in \cite{Frigo-etAl2017}. Since the acquired signal is affected by simultaneous dynamic trends, PMU estimates might exhibit abrupt changes that produce amplitude and phase steps in the recovered waveform. Furthermore, the fundamental component parameters are updated every 20 ms, in accordance with the PMU reporting rate of 50 fps.
	
	In order to obtain a smoother and more realistic trend, we develop a non-linear model that reproduce the acquired signal fluctuations through the combined effect of test conditions provided by IEEE Std, and is defined as follows:
	\begin{eqnarray}
		\label{eq:SiL_model}
		y(t) =&& A \cdot (1 + k_A\cdot \cos(2\pi f_A t)) \cdot \\ 
		&&\cos(2\pi f t + \varphi + k_\varphi \cdot \cos(2\pi f_\varphi t) + R_f t^2) \nonumber
	\end{eqnarray}
	The non-linear model parameters are summarized in Tab. \ref{tab:SiL_mod_par}. 
	We approximate amplitude and phase fluctuations through modulation terms compliant with IEEE Std measurement bandwidth test. In particular, the amplitude modulation term is characterized by a depth $ k_A $ and frequency $ f_A $, equal to 13.6\% and 153.1 mHz, respectively. In a similar way, the phase modulation term is characterized by a depth $ k_\varphi $ and frequency $ f_\varphi $, equal to 56.4 mrad and 152.6 mHz.
	
\begin{figure}
		\centering
		\includegraphics[width=.5\columnwidth]{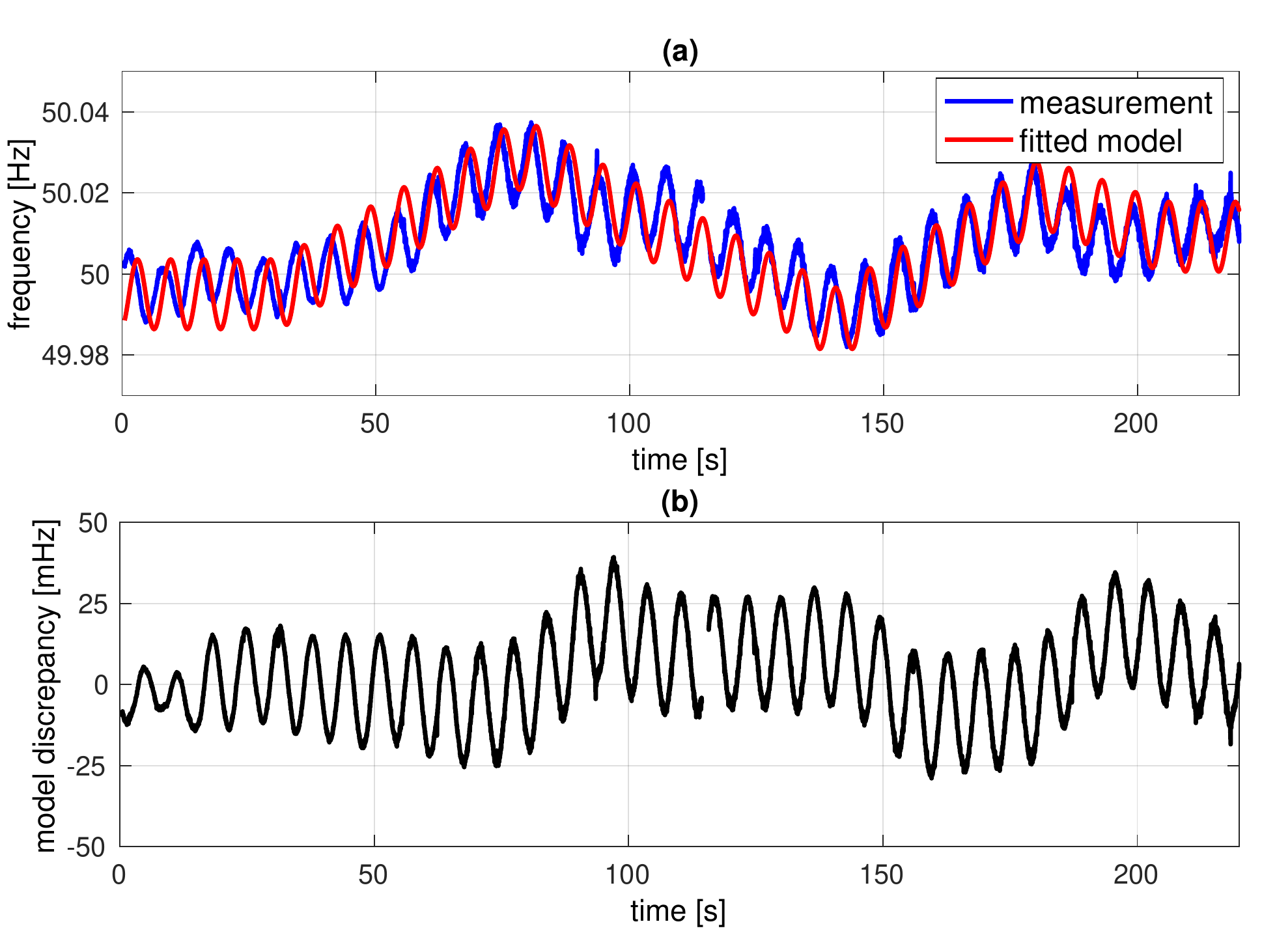}
		\caption{In (a), comparison between frequency raw measurement and the corresponding nonlinear fitted model, related to the large inter-area oscillation event reported in \cite{SiL-report}. In (b), discrepancy between fitted model and measurement.}
		\label{fig:SiL_freq}
	\end{figure}

	Frequency increasing and decreasing trends are modeled as a sequence of positive and negative linear ramps compliant with IEEE Std frequency ramp test. For the sake of completeness, Tab. \ref{tab:SiL_rocof} reports the time-duration of each waveform segment, and the corresponding ramp parameter $ R_f $, whose value varies within the restricted range [-2.30, 2.55] mHz/s.
	In this regard, Fig. \ref{fig:SiL_freq}(a) compares the frequency values measured by the PMU (blue) with the corresponding non-linear fitted model (red). It is worth noticing that the proposed approach enables us to limit the discrepancy between measurement data and mathematical model, and at the same time to guarantee a smooth trend of fundamental component (Fig. \ref{fig:SiL_freq}(b)). 
	Finally, we reproduce a more realistic measurement noise scenario by adding a white Gaussian noise with SNR equal to 60 dB.
	
	\begin{table}
		\centering
		\caption{Dataset II: Non-Linear Fit Model Parameters}
		\label{tab:SiL_mod_par}
		\begin{tabular}{ccccccc}
			\toprule
			$ A $ & $ f $ & $ \varphi $ & $ k_A $ & $ f_A $& $ k_\varphi $& $ f_\varphi $ \\
			$ [$kV$] $ & $ [ $Hz$ ] $ & $ [ $rad$ ] $ & $ [$\%$] $ & $ [$mHz$] $ & $ [$mrad$] $ & $ [$mHz$] $ \\
			\midrule
			71.45 & 50.02 & -1.80 & 13.6 & 153.1 & 56.4 & 152.6 \\
			 \bottomrule
		\end{tabular}
	\end{table}

		\begin{table}
		\centering
		\caption{Dataset II: Non-Linear Fit Ramp Parameter Values}
		\label{tab:SiL_rocof}
		\begin{tabular}{c|ccccccc}
			\toprule
			\textbf{Segment} &\textbf{1} &\textbf{2} &\textbf{3} &\textbf{4} &\textbf{5} &\textbf{6} &\textbf{7} \\
			\midrule
			$ t_{start} [$s$ ] $ & 0 & 30.5 & 78.5 & 98.5 & 140.5 & 180.5 & 204.5 \\
			$ t_{stop} [$s$ ] $ & 30.5 & 78.5 & 98.5 & 140.5 & 180.5 & 204.5 & 220.5 \\
			$ R_f [$mHz/s$ ] $ & - & 2.28 & -2.29 & -2.05 & 2.53 & -1.42 & - \\ 
			\bottomrule
		\end{tabular}
	\end{table}
	
	Based on model \eqref{eq:SiL_model} and parameters of Tab. \ref{tab:SiL_mod_par}, we are able to define the ground-truth ROCOF at each sampling time. With respect to this reference value, we thoroughly characterize the ROCOF estimation accuracy of csTFM, eIpDFT and iIpDFT. 
	
\begin{figure}
		\centering
		\includegraphics[width=.5\columnwidth]{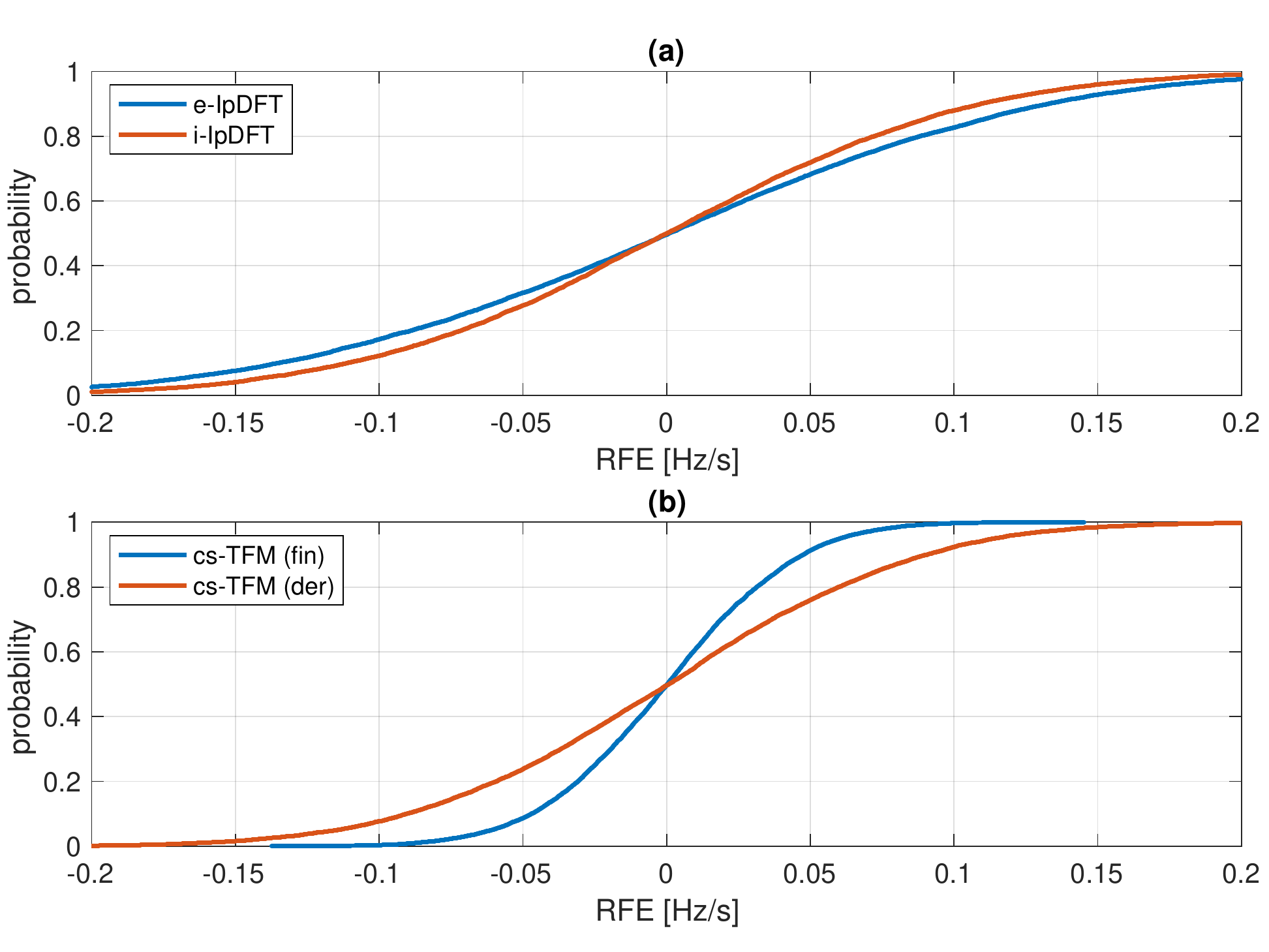}
		\caption{Given a window length of 60 ms (P-class configuration), comparison of the cumulative distribution functions associated to the RFEs as produced by static (a) and dynamic model-based algorithms (b).}
		\label{fig:sil_cdf_3c}
	\end{figure}

	In Fig. \ref{fig:sil_cdf_3c} we present the error CDFs associated to the P-class configuration for static (a) and dynamic (b) approaches. It is interesting to observe that the static approaches as well as the derivative formulation of cs-TFM provide a nearly coincident performance, with an estimation error ranging within $ \pm $0.2 Hz/s. Conversely, the finite-difference formulation of cs-TFM enables us to further reduce the maximum estimation error up to 0.1 Hz/s. As reported in Table \ref{tab:sil_result}, though, none of the considered estimation techniques provides a sufficient correlation with the ground-truth values.
	
	\begin{table}
		\centering
		\caption{Dataset II: RFE Statistical Properties}
		\label{tab:sil_result}
		\begin{tabular}{c|c||ccc|c}
				\toprule
				\textbf{Estimator} & \textbf{Class} & \textbf{Mean} & \textbf{Std Dev} & \textbf{95\%} & \textbf{Corr}\\
				\midrule
				\multirow{2}{*}{e-IpDFT} & P & 2.51$ \cdot $10$ ^{-6} $ & 0.10 & 0.24 & 5.65\% \\
				& M & 2.80$ \cdot $10$ ^{-6} $ & 0.04 & 0.09 & 14.58\%\\
				\midrule
				\multirow{2}{*}{i-IpDFT} & P & -4.48$ \cdot $10$ ^{-6} $ & 0.09 & 0.20 & 6.80\%\\
				& M & -3.24$ \cdot $10$ ^{-6} $ & 0.03 & 0.07 & 18.24\% \\
				\midrule
				cs-TFM & P & 2.07$ \cdot $10$ ^{-6} $ & 0.04 & 0.09 & 15.64\%\\
				(fin) & M & -1.68$ \cdot $10$ ^{-6} $ & 0.01 & 0.03 & 39.56\% \\
				\midrule
				cs-TFM & P & 9.68$ \cdot $10$ ^{-5} $ & 0.07 & 0.16 & 8.38\%\\
				(der) & M & -3.00$ \cdot $10$ ^{-6} $ & 0.02 & 0.04 & 29.19\% \\
				\bottomrule
		\end{tabular}
	\end{table}

\begin{figure}
		\centering
		\includegraphics[width=.5\columnwidth]{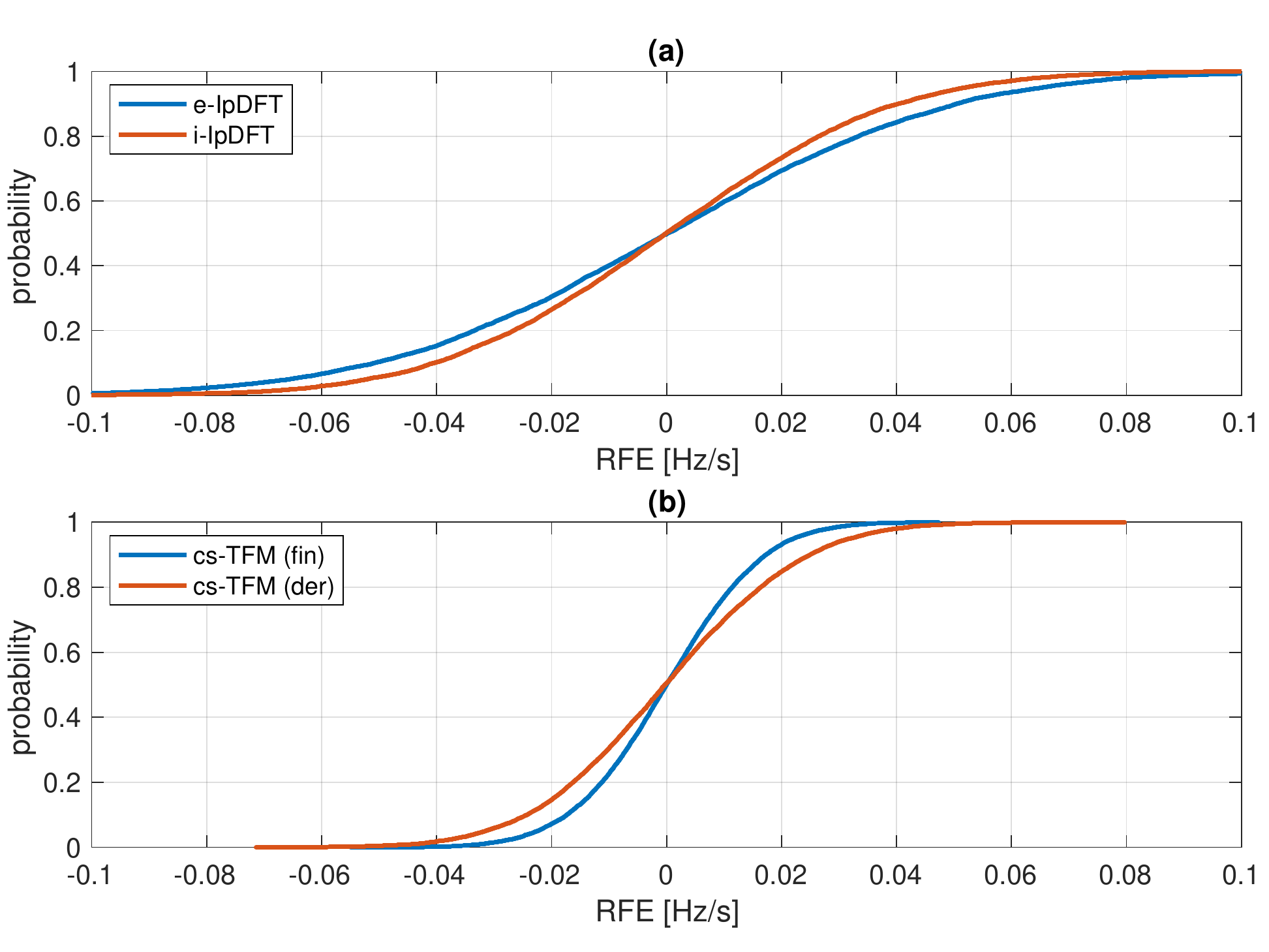}
		\caption{Given a window length of 100 ms (M-class configuration), comparison of the cumulative distribution functions associated to the RFEs as produced by static (a) and dynamic model-based algorithms (b).}
		\label{fig:sil_cdf_5c}
	\end{figure}

	Fig. \ref{fig:sil_cdf_5c} represents the error CDFs associated to the M-class configuration for static (a) and dynamic (b) approaches. Differently from the previous dataset, the increased window length does not produce a remarkable performance enhancement. As regards the static approaches, the RFE variation range is almost halved, with a maximum estimation error of 0.1 Hz/s. The dynamic approaches, instead, provide a similar performance with RFE limited within $ \pm $0.05 Hz/s. Once more, the finite difference formulation provides the most accurate solution, exploiting both the dynamic signal model and the filter smoothing effect. However, despite this error reduction, the estimated values prove to be very poorly correlated with the ground-truth reference (just 39.56\% in the best case).

	A motivation could be found in the peculiar variational trend of the fundamental frequency. Both oscillatory and linear trends are characterized by time constants in the order of tens of seconds. The strict requirements in terms of reporting rate and latency force the PMUs to consider short observation periods where these phenomena can be hardly discriminated from steady-state conditions. In this sense, the employment of canonical PMUs for similar scenarios should probably require a close analysis of the most suitable estimation approaches, observation window lengths and reporting rates.
	
\begin{figure}
		\centering
		\includegraphics[width=.5\columnwidth]{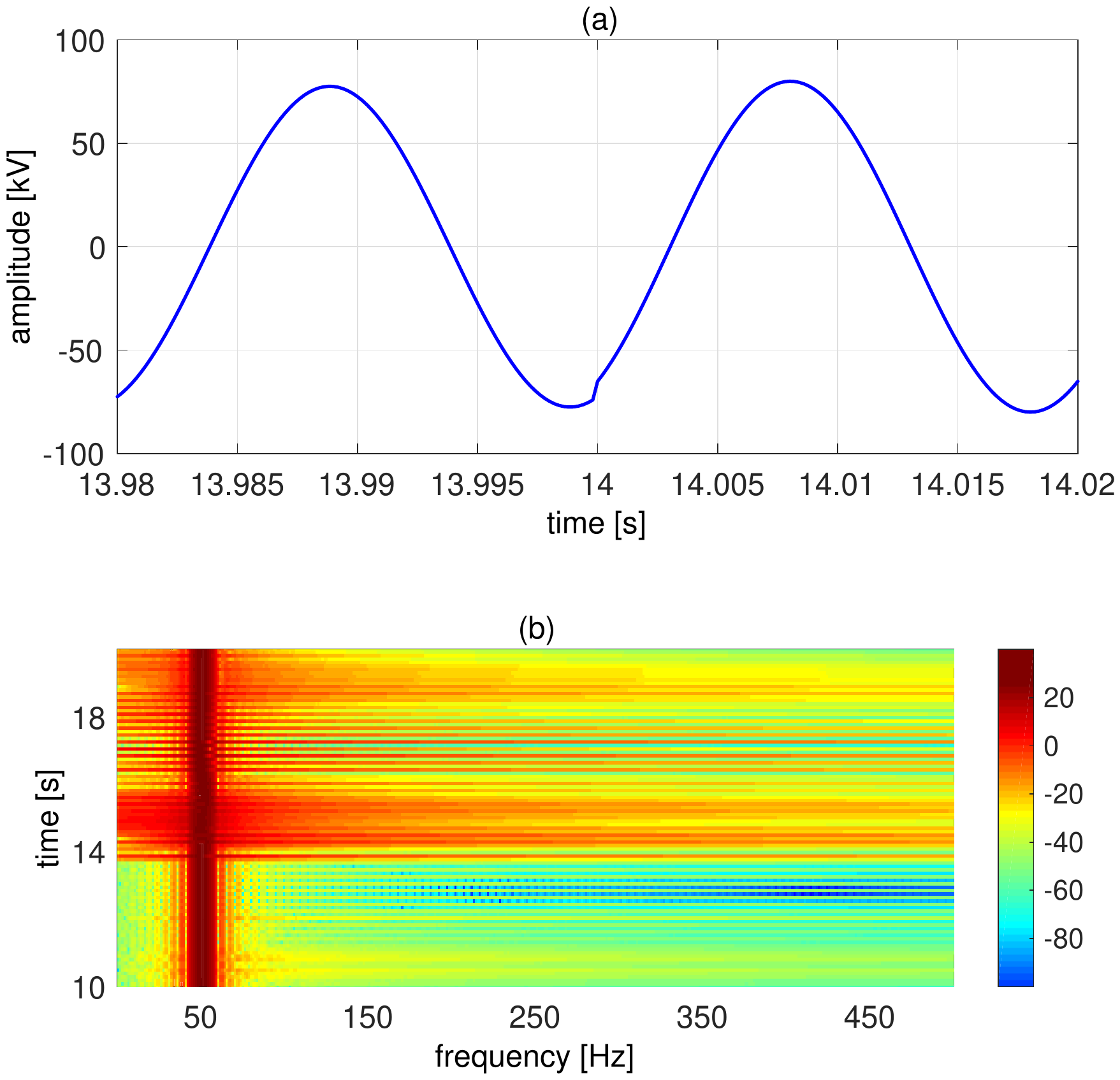}
		\caption{In (a), time-domain acquisition in correspondence of the islanding operation at 14 s. In (b), the corresponding spectrogram expressed in dB.}
		\label{fig:isl_time}
	\end{figure}

	\paragraph{Dataset III} The third dataset consists of the waveforms acquired by an oscilloscope during an intentional islanding maneuver of an urban medium voltage power network, carried out in Imola, Italy, on August 13, 2009 \cite{Borghetti-etAl2011}. In this regard, Fig. \ref{fig:isl_time}(a) represents the waveform portion of interest, where the islanding operation occurs around 14 s, whereas Fig. \ref{fig:isl_time}(b) shows the spectrogram (i.e. Short Time Fourier Transform as function of time) computed over windows of 1024 samples. Before the islanding maneuver, the signal consists in the fundamental tone only. Once the operation has started, instead, the signal energy is no more limited around the fundamental tone, but spreads among over 500 Hz. In such conditions, the nominal PMU pass-band is exceeded and even the synchrophasor definition, relying on a finite spectrum assumption, lacks of appropriateness.
	
	\begin{table}
		\centering
		\caption{Dataset III: Fitted Waveform Parameters}
		\label{tab:wavpar}
		\begin{tabular}{c|ccc}
			\toprule
			 \textbf{Segment} & \textbf{Amplitude [kV]} & \textbf{Frequency [Hz]} & \textbf{Phase [rad]}\\
			 \midrule
			 pre-islanding & \multirow{2}{*}{77.66} & \multirow{2}{*}{50.07} & \multirow{2}{*}{1.032}\\
			 ($ t < $ 14 s) & & & \\
			 \midrule
			 post-islanding & \multirow{2}{*}{80.14} & \multirow{2}{*}{50.07} & \multirow{2}{*}{0.884}\\
			 ($ t \geq $ 14 s) & & & \\
			 \midrule
			 \midrule
			 \multicolumn{4}{c}{SNR = 46.24 dB, \; THD = 2.31\%, \; SINAD = 33.49 dB}\\
			 \bottomrule
		\end{tabular}
	\end{table}

	In this case, we do not have any direct measurement or \textit{a priori} information regarding the frequency reference value. In order to extract the waveform parameters before and after the islanding maneuver, we apply the non-linear fitting routine presented in \cite{Frigo-etAl2017}. Given the signal model of IEEE Std \textit{Step change} test, we are able to detect a quasi-instantaneous variation of 3.19\% and 0.148 rad for amplitude and phase, respectively, as given by the waveform parameters reported in Table \ref{tab:wavpar}. In Matlab, we also quantify the noise and distortion level by computing the SNR, THD and SINAD indexes, that are equal to 46.24 dB, 2.03\% and 33.49 dB, respectively.
	
	Since the adopted signal model does not account for frequency variations, the corresponding ROCOF reference value is fixed at 0 Hz/s. In this scenario, the performance evaluation based only on the RFE assessment does not provide any information regarding the estimator behavior during the transient. Therefore, we consider a time-based metric, introduced and further discussed in \cite{Frigo-etAl2017}, i.e. the normalized root-mean-squared error (nRMSE). At each reporting time, the synchrophasor estimation algorithms provide an updated estimate of amplitude, phase, frequency and ROCOF. Based on this information, we recover the fundamental component trend in the time domain, and compute the root-mean-squared error with respect to the sample window considered during the estimation process. Finally, we normalize the obtained RMSE by the sample window energy. In this way, the new metric accounts for the portion of signal energy that has been neglected or misrepresented by algorithm estimates.
	
	In real test conditions, the nRMSE value depends on many variables, like the signal model and parameters, the noise and distortion level, and the algorithm estimation error. As a consequence, it is not possible to define a universal criterion of measurement reliability. Nevertheless, in a controlled scenario, it is possible to characterize the distribution of nRMSE values in the expected operating conditions and thus set a suitable threshold for transient detection. In the present case, based on the parameters reported in Table \ref{tab:wavpar}, it is possible to infer the expected distribution of nRMSE value before and after the islanding maneuver. In Matlab, we reproduce a test waveforms with the same model parameters and distortion levels. Over a test duration of 1 s, we compute the corresponding nRMSE as provided by the three algorithms. In Table \ref{tab:nrmse}, we report the statistical description of the nRMSE distribution in terms of mean, standard deviation and maximum value. It is interesting to observe how these values nearly coincide with the experimental results. Based on this preliminary characterization, it is thus possible to suitably define a detection threshold that is triggered only by large and unexpected variations.
	
	\begin{table}
		\centering
		\caption{Dataset III: nRMSE Statistical Distribution}
		\label{tab:nrmse}
		\begin{tabular}{c|c||ccc}
			\toprule
			\multirow{2}{*}{\textbf{Segment}} & \multirow{2}{*}{\textbf{Class}} & \textbf{Mean} & \textbf{Std Dev} & \textbf{Max} \\
			& & \textbf{[ppm]} & \textbf{[ppm]} & \textbf{[ppm]} \\
			\midrule
			pre-islanding & P & 81.94 & 1.34 & 85.70\\
			($ t < $ 14 s) & M & 52.40 & 0.77 & 54.31\\
			\midrule
			post-islanding & P & 48.89  & 1.35  & 52.94\\
			($ t \geq$ 14 s) & M & 29.98 & 0.69 & 32.07\\
			\bottomrule
		\end{tabular}
	\end{table}

\begin{figure}
		\centering
		\includegraphics[width=.5\columnwidth]{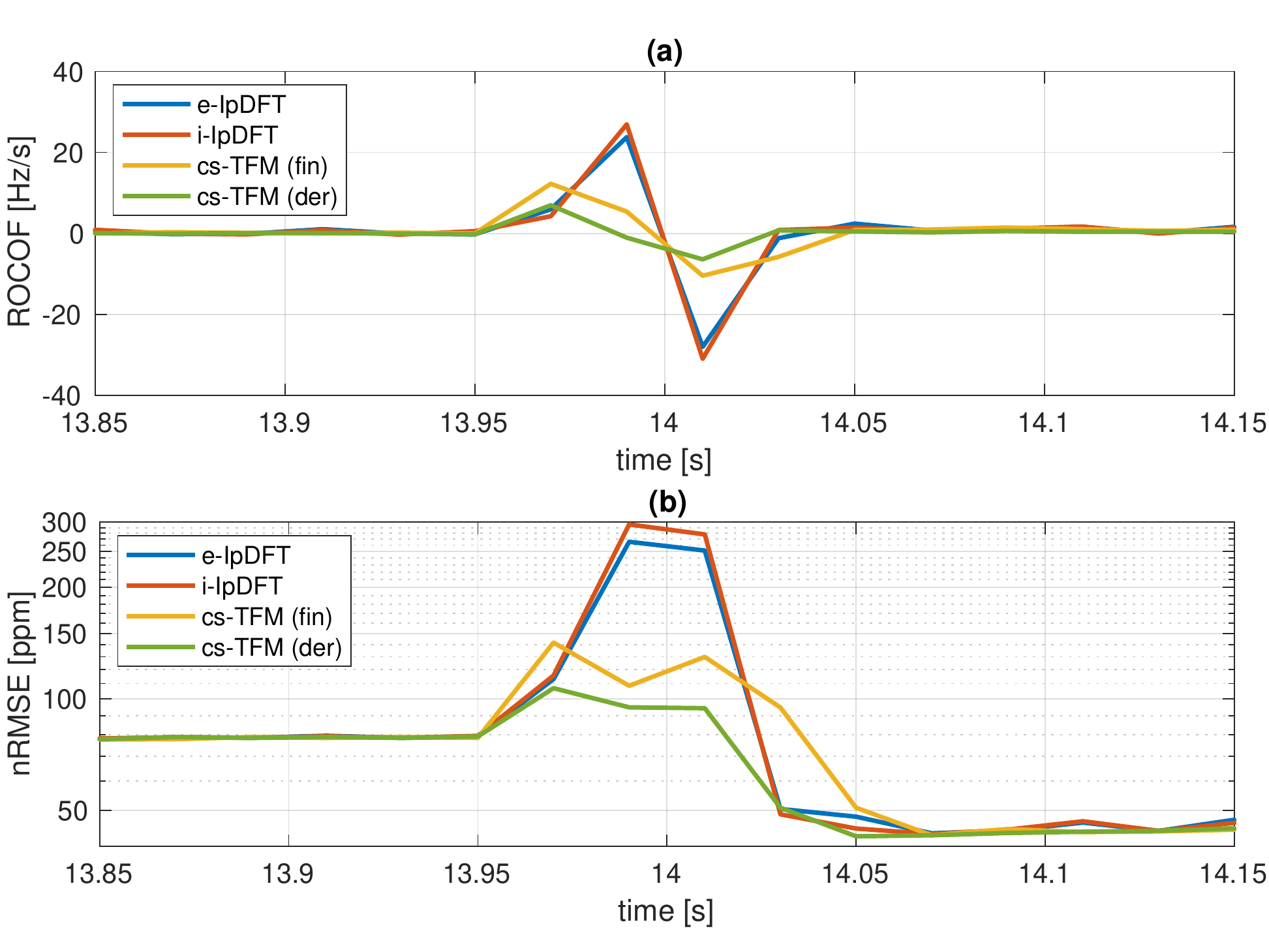}
		\caption{In P-class configuration (window length of 60 ms), ROCOF estimates as measured during the islanding maneuver (a), and estimation uncertainty as quantified by nRMSE (b).}
		\label{fig:imola_rmse_3c}
	\end{figure}

	In Fig. \ref{fig:imola_rmse_3c}(a) we report the ROCOF estimates in three-cycle configuration, in correspondence of the islanding maneuver. Fig. \ref{fig:imola_rmse_3c}(b) presents the corresponding normalized nRMSEs that can be considered as a measure of algorithms' reliability. Both static and dynamic ROCOF estimates exhibit a rapid and significant increase, up to maximum values around 20 Hz/s. Nevertheless, the similar increase of normalized RMSE suggests that a significant portion of signal energy has been neglected or misrepresented, e.g. in the presence of high distortion levels or transient events. In the latter case, the finite spectrum assumption is not satisfied and the PMU estimation accuracy is not guaranteed anymore. 
	
\begin{figure}
		\centering
		\includegraphics[width=.5\columnwidth]{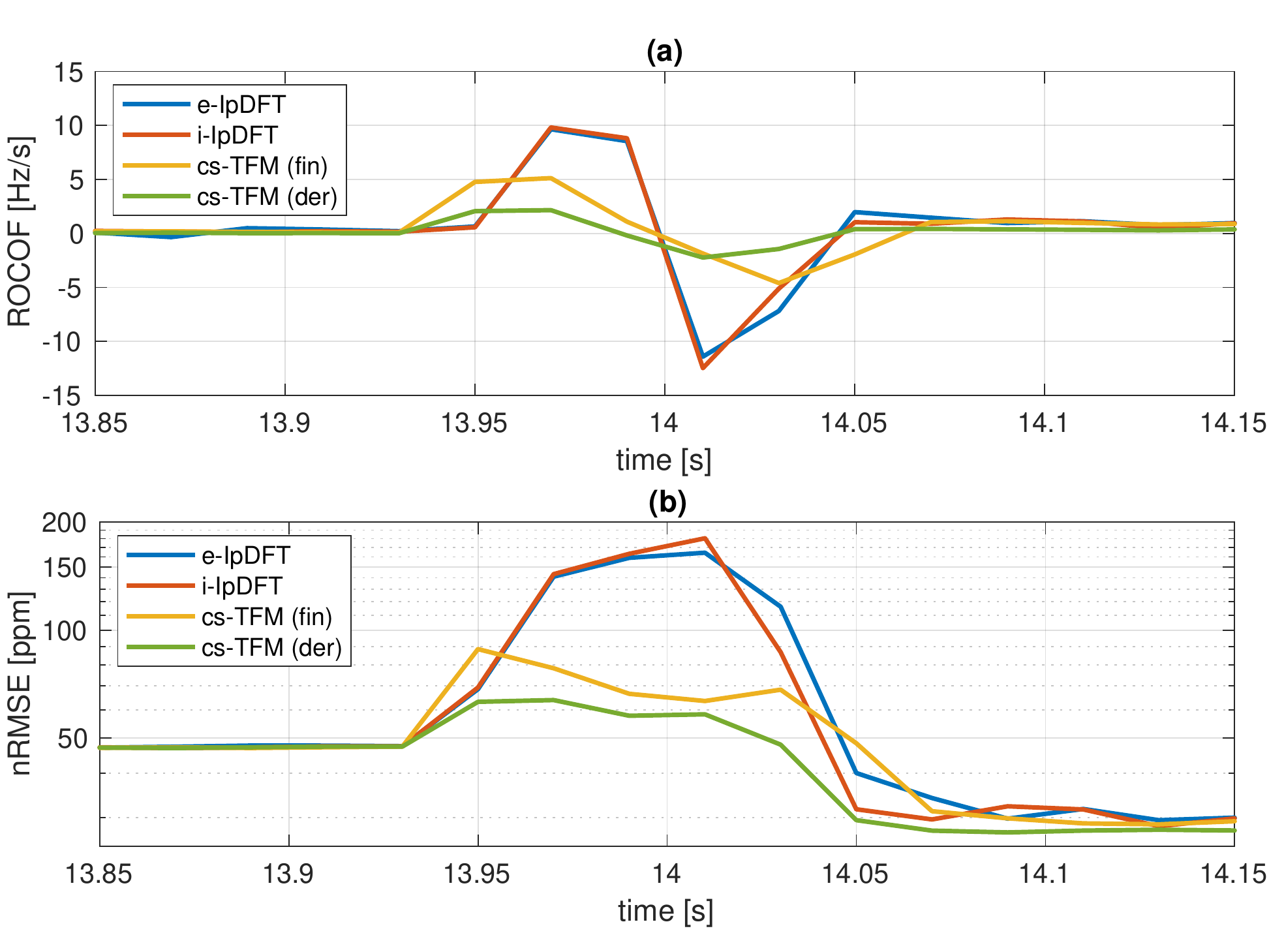}
		\caption{In M-class configuration (window length of 100 ms), ROCOF estimates as measured during the islanding maneuver (a), and estimation uncertainty as quantified by nRMSE (b).}
		\label{fig:imola_rmse_5c}
	\end{figure}

	Similar considerations can be made for the five-cycle configuration in Fig. \ref{fig:imola_rmse_5c}(a). A longer window allows for a more accurate ROCOF estimation, though still totally unreliable as shown by the large nRMSE values in Fig. \ref{fig:imola_rmse_5c}(b).
	
	Given the synchrophasor model inconsistency in non-stationary conditions, it is reasonable to expect high RFE during transient events. For this reason, many practical implementation of PMU-based ROCOF meters provide a \textit{blocking} scheme that prevents unrealistic ROCOF values to be propagated to the control schemes. Once detected a transient event (e.g. by means of a metric similar to nRMSE), one of the most common solutions consists in fixing the ROCOF estimate to the last \textit{valid} value, i.e. to the last estimate before the event occurrence. In this paper, though, such mechanism is not implemented as the proposed analysis aims at assessing the ROCOF estimation accuracy. In this context, the evaluation of the actual behavior during a transient event might provide interesting insights for the design and development of more accurate and robust estimators.
	
	\paragraph{Result discussion}
	As a performance summary, Table \ref{tab:res_summary} collects the estimation accuracy limits provided by each ROCOF estimation technique in the three considered datasets. In this context, we compare the obtained results with the most severe requirements of the IEEE Std tests \cite{ieeeC37,ieee_am} that better approximate the considered operating scenarios.
	
	\begin{table*}
		\centering
		\caption{ROCOF Estimation Accuracy vs IEEE Std Requirements}
		\label{tab:res_summary}
		\begin{tabular}{c||c|c||cccc|c}
				\toprule
				\textbf{Performance} & \textbf{Test} & \textbf{Window} & \textbf{e-IpDFT} & \textbf{i-IpDFT} & \textbf{cs-TFM} & \textbf{cs-TFM} &  \textbf{IEEE Std.} \\
				\textbf{Metric} & \textbf{Waveform} & \textbf{Length} & \textbf{(finite)} & \textbf{(finite)} & \textbf{(finite)} & \textbf{(derivative)}  & \textbf{Limit} \\
				\midrule
				\multirow{4}{*}{RFE [Hz/s]} & Hydro-Quebec & class P - 60 ms & 10.51 & 5.59 & 1.12 & 1.11  & 0.4\\
				& Power System & class M - 100 ms & 2.03 & 0.57 & 0.38 & 0.56 & suspended\\
				\cmidrule{2-8}
				& Inter-Area & class P - 60 ms & 0.24 & 0.20 & 0.09 & 0.16 & 0.4 \\
				& Oscillation & class M - 100 ms & 0.09 & 0.07 & 0.03 & 0.04 & 0.2 \\
				\midrule
				\multirow{2}{*}{nRMSE [ppm]} & Islanding & class P - 60 ms & 263 & 297 & 112 & 146 &  - \\
				& Maneuver & class M - 100 ms & 161 & 180 & 62 & 84 &  - \\
				\bottomrule
		\end{tabular}
	\end{table*}

	As regards the Hydro-Québec grid, the term of comparison is given by the \textit{Harmonic distortion} test, whose maximum RFE is set equal to 0.4 Hz/s for P-class, whereas it is currently suspended for M-class. Given a window length of 60 ms, none of the considered algorithm proves to be compliant with the IEEE Std requirement. Nevertheless, it should be noticed that the simulated scenario represents a much more challenging test-bench as it includes several narrow-band disturbances as well as fundamental time-varying parameters. In M-class configuration, the increased window length produces a significant performance enhancement for the algorithms capable of mitigating out-of-band disturbances (i-IpDFT) and accounting for dynamic trends (cs-TFM).
	
	The inter-area oscillation is compared to the \textit{Frequency ramp} test, that requires RFE not to exceed 0.4 and 0.2 Hz/s for P- and M-class, respectively. In this case, all the estimation techniques satisfy these requirements for both window lengths. As previously discussed, though, the achieved accuracy level is not sufficient to suitably track the ROCOF long-term oscillations. Before being deployed in such a scenario, PMUs are required to improve their resolution power, i.e. their capability to accurately detect even reduced variations of ROCOF as typical of normal operating conditions. On the other hand, it should be noticed that the simulated condition is further complicated by the additive measurement noise, whose level is comparable with the quantities to be measured.
	
	For the islanding maneuver, the closest IEEE Std test condition is the \textit{Amplitude} or \textit{Phase step change}. However, none of these tests define any limit regarding the ROCOF overshoot, but only a maximum response time equal to 120 and 280 ms for P- and M-class, respectively. Within these time windows, the PMU-based measurements are allowed to vary without any correlation with precedent or subsequent ROCOF values \cite{Roscoe-etAl2017}. Indeed, during transient events, the signal spectrum is continuous and cannot be represented by the synchrophasor signal model conventionally employed by window-based techniques. Conversely, the proposed performance metric, nRMSE, accounts for the amount of energy that is discarded by the estimated solution. In general, it can be interpreted as a measure of the estimates reliability, but it is worth noticing that the entire synchrophasor representation looses its mathematical consistency during a transient event.
	
	Based on the obtained results, it is possible to infer some recommendations for PMU-based ROCOF measurements. First, the employment of a dynamic signal model allows for a more accurate tracking of time-varying trends, like linear ramps or oscillations. On the other hand, a direct computation of the frequency first time-derivative proves to be gravely affected by spurious injections from measurement noise or uncompensated disturbances. Therefore, it is preferable to adopt a finite difference formulation that produces a smoothing effect on the final ROCOF estimates, at the cost of a slightly performance deterioration in the presence of transient events.
	
	\section{Under-Frequency Load-Shedding Application}
\begin{figure}
		\centering
		\includegraphics[width=.5\columnwidth]{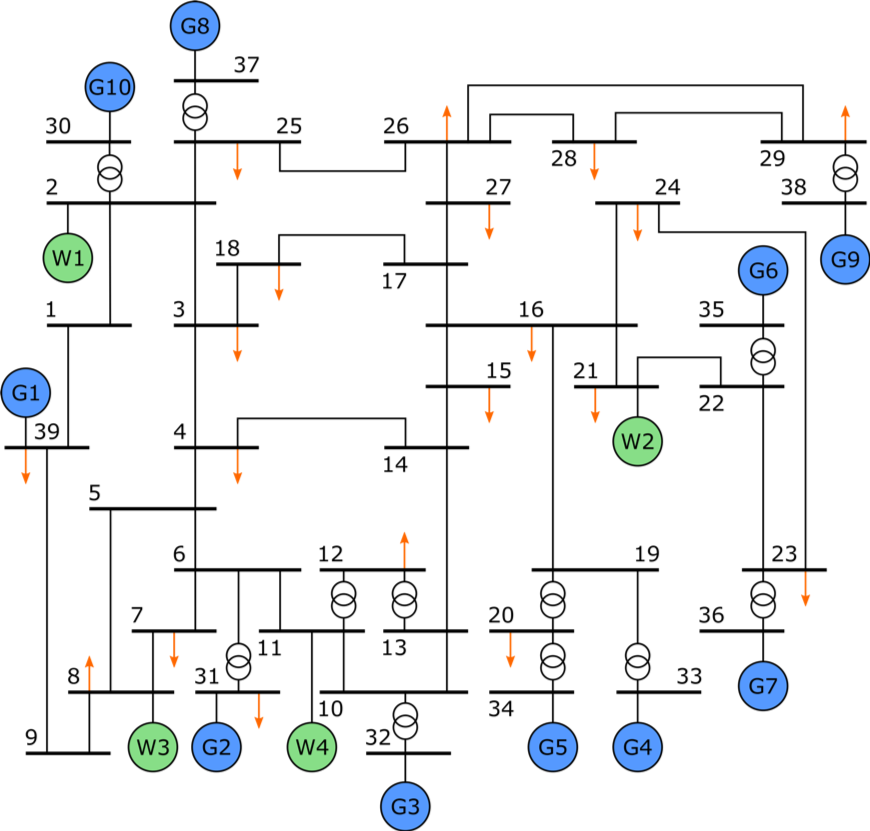}
		\caption{Simulated network configuration with 10 synchronous generators (blue circles), 4 wind farms (green circles), and 19 loads (orange arrows).}
		\label{fig:ieee39_grid}
	\end{figure}

	In this Section, we test the actual feasibility of PMU-based ROCOF measurements by considering a realistic scenario of Under-Frequency Load-Shedding application \cite{Derviskadic-etAl2018}. For this analysis, we adopt the Opal-RT eMEGAsim PowerGrid Real Time Simulator \cite{OpalRT} to implement a suitably modified dynamic model of the IEEE 39-bus power system \cite{github}, that is extensively illustrated in \cite{Derviskadic-etAl2018} and displayed in Fig. \ref{fig:ieee39_grid}.
	
	The grid consists of 39 buses, 10 conventional generators, and 19 loads, with a nominal voltage of 345 kV. The inclusion of 4 wind farms accounts for the effects of large-scale renewable generation. Both generation and load profiles are not constant, but inferred from real-world measurements \cite{Zuo-etAl2018}. In order to account for the dynamic behavior of the loads, the EPRI LOADSYN model has been adopted \cite{load_model}. 
	
\begin{figure}
		\centering
		\includegraphics[width=.5\columnwidth]{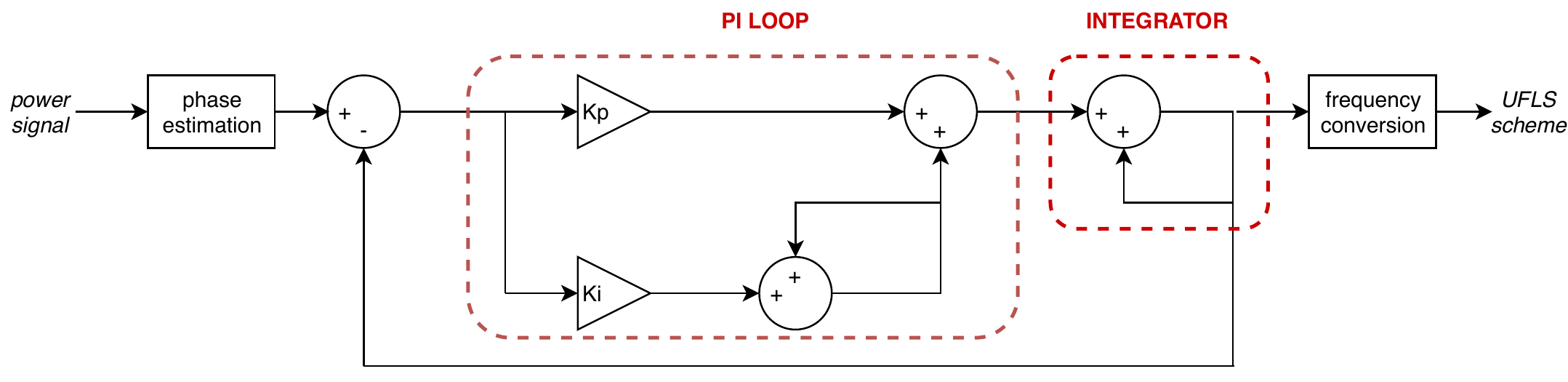}
		\caption{Block scheme of the adopted PLL-based frequency estimator.}
		\label{fig:pll-scheme}
	\end{figure}

	In each load bus, we alternatively place a Phase-Locked Loop (PLL) or two PMUs, in the following referred to as PMU-1 and PMU-2. In particular, the PLL (as implemented in the Matlab library \cite{PLL}, according to the block scheme in Fig. \ref{fig:pll-scheme}) is characterized by a proportional and integral gain equal to 180 and 3200, respectively, PMU-1 is a P-class PMU that relies on a static signal model, and PMU-2 is a M-class PMU that is based on a dynamic signal model and computes ROCOF as the instantaneous time-derivative of frequency. In the real-time simulation, we develop and compare two different UFLS schemes: one based on the frequency estimate provided by the PLL, and the other based on the ROCOF estimates provided by the PMUs. The frequency-based scheme is directly inferred by ENTSO-E guidelines \cite{ENTSO-E}, whereas the ROCOF-based scheme has been designed with specific reference to the low-inertia properties of the simulated grid (further details in \cite{Derviskadic-etAl2018}).
	
	Within the simulated scenario, we evaluate two operational aspects. On one side, we investigate whether the employment of PMU-based ROCOF relays provides a significant improvement with respect to traditional frequency-based approaches. On the other side, we experimentally validate the sensitivity analysis of Section IV. In particular, we assess whether the performance class (i.e. the window length) and the signal model (i.e. static or dynamic) affect significantly the UFLS scheme outcomes.
	
\begin{figure}
		\centering
		\includegraphics[width=.5\columnwidth]{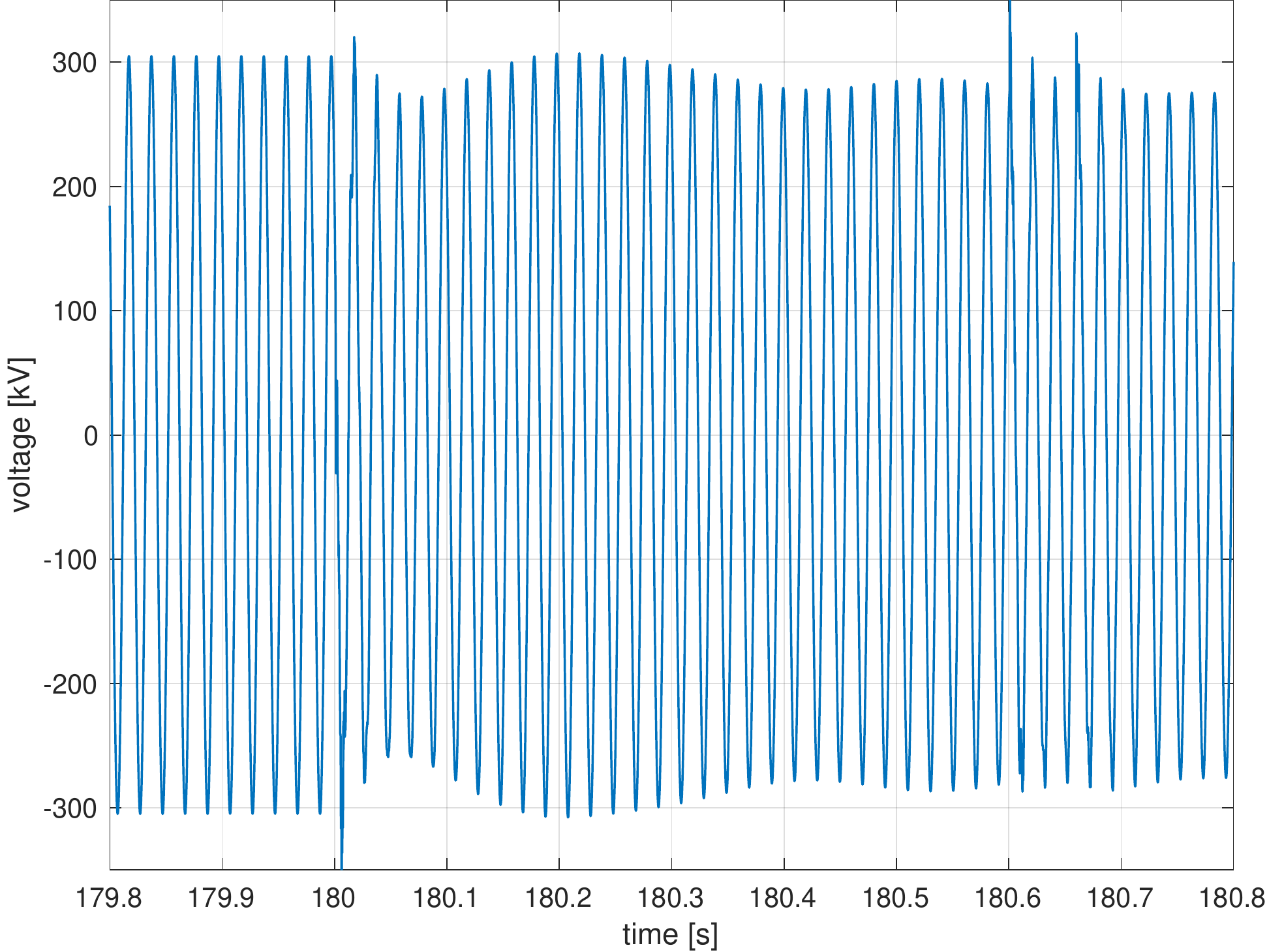}
		\caption{Time-domain trend of the voltage waveform acquired by the PMUs in the node 23 (SNR = 80 dB).}
		\label{fig:ieee39_time}
	\end{figure}

	For this analysis, we simulate the outage of three generators (G4, G6 and G7) for a total tripped power of 1.5 GW. Fig. \ref{fig:ieee39_time} presents the corresponding raw voltage waveform as acquired by the PMUs placed in the node 23. Such a contingency might result in an unavoidable blackout, in case the UFLS does not provide an adequate countermeasure in terms of response time and amount of shed loads. Compared to the frequency-based solutions, the predictive effect of ROCOF measurements might allow for a prompt and decisive and response.
	
	The outage occurs at $ t $ = 180 s and produces a noticeable step change that is followed by a long-term oscillating trend due to the grid inertia properties and the gradual load shedding and reconnection. In order to model a plausible level of measurement noise, the acquired waveform is intentionally corrupted by an additive and uncorrelated white Gaussian noise (SNR = 80 dB). As a consequence, the ROCOF estimation techniques face two main challenges: on one side, the non-stationary conditions with both fast transients and slow modulations, on the other side, the wide-band additive noise.
	
\begin{figure}
		\centering
		\includegraphics[width=.5\columnwidth]{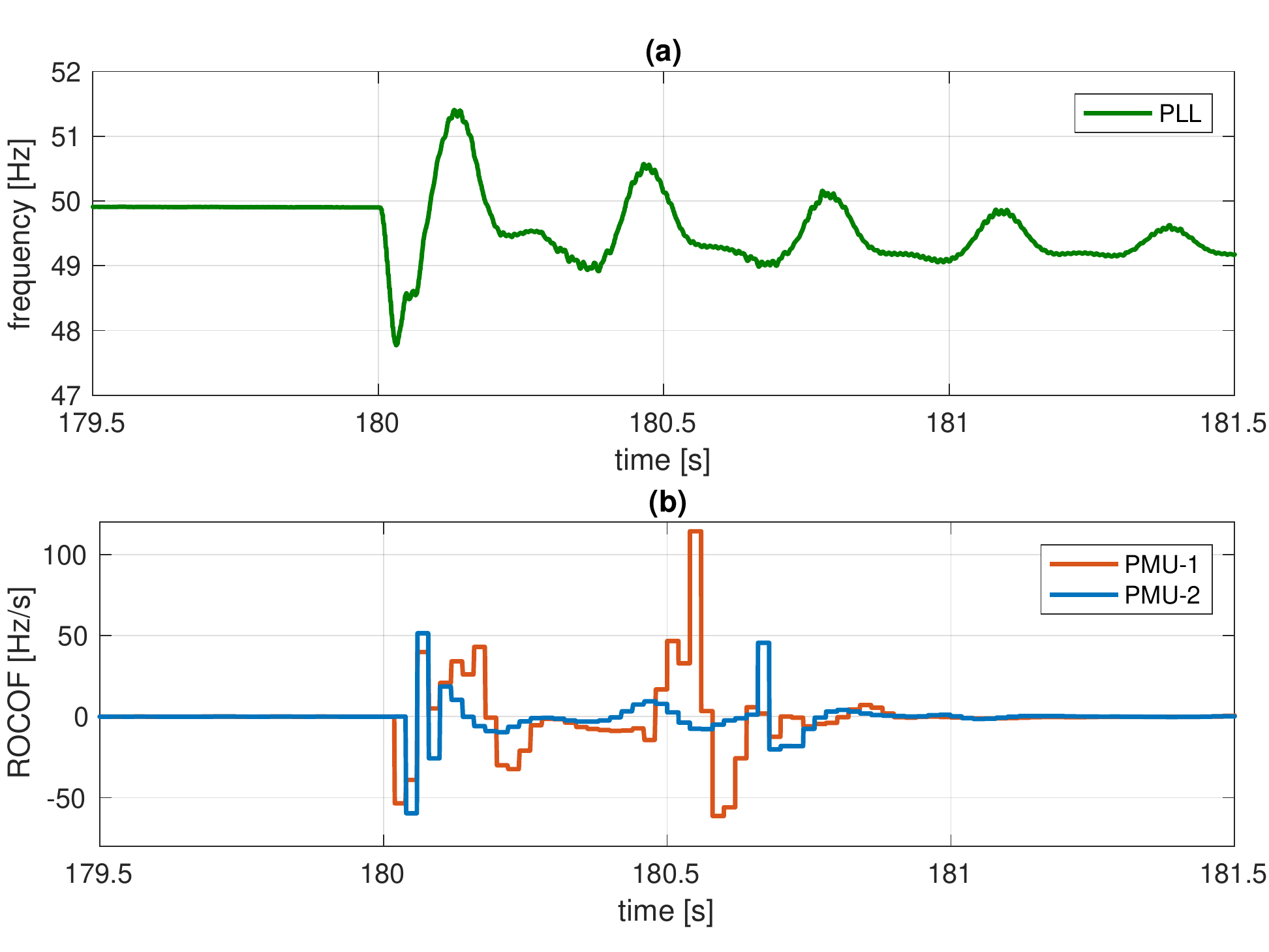}
		\caption{Profile of frequency (a) and ROCOF (b) as measured at node 23 by PLL, PMU-1 and PMU-2 in green, red and blue line, respectively.}
		\label{fig:ufls_rocof}
	\end{figure}

	Fig. \ref{fig:ufls_rocof} presents the evolution of the control variables for the considered UFLS schemes in correspondence of the outage occurrence. In the upper graph, the frequency estimated by the PLL is characterized by a sequence of dampened oscillations, whereas the lower graph\footnote{It should be noticed that the ROCOF estimates in Fig. \ref{fig:ufls_rocof}(b) are discrete-time, since both PMUs adopt a reporting rate of 50 fps.} shows the ROCOF estimated by PMU-1 and PMU-2. In this regard, it is worth observing that the estimates of the two PMUs differ significantly in terms of both absolute value and polarity, in confirmation of the algorithm-dependency of PMU-based ROCOF measurements.
	
\begin{figure}
		\centering
		\includegraphics[width=.5\columnwidth]{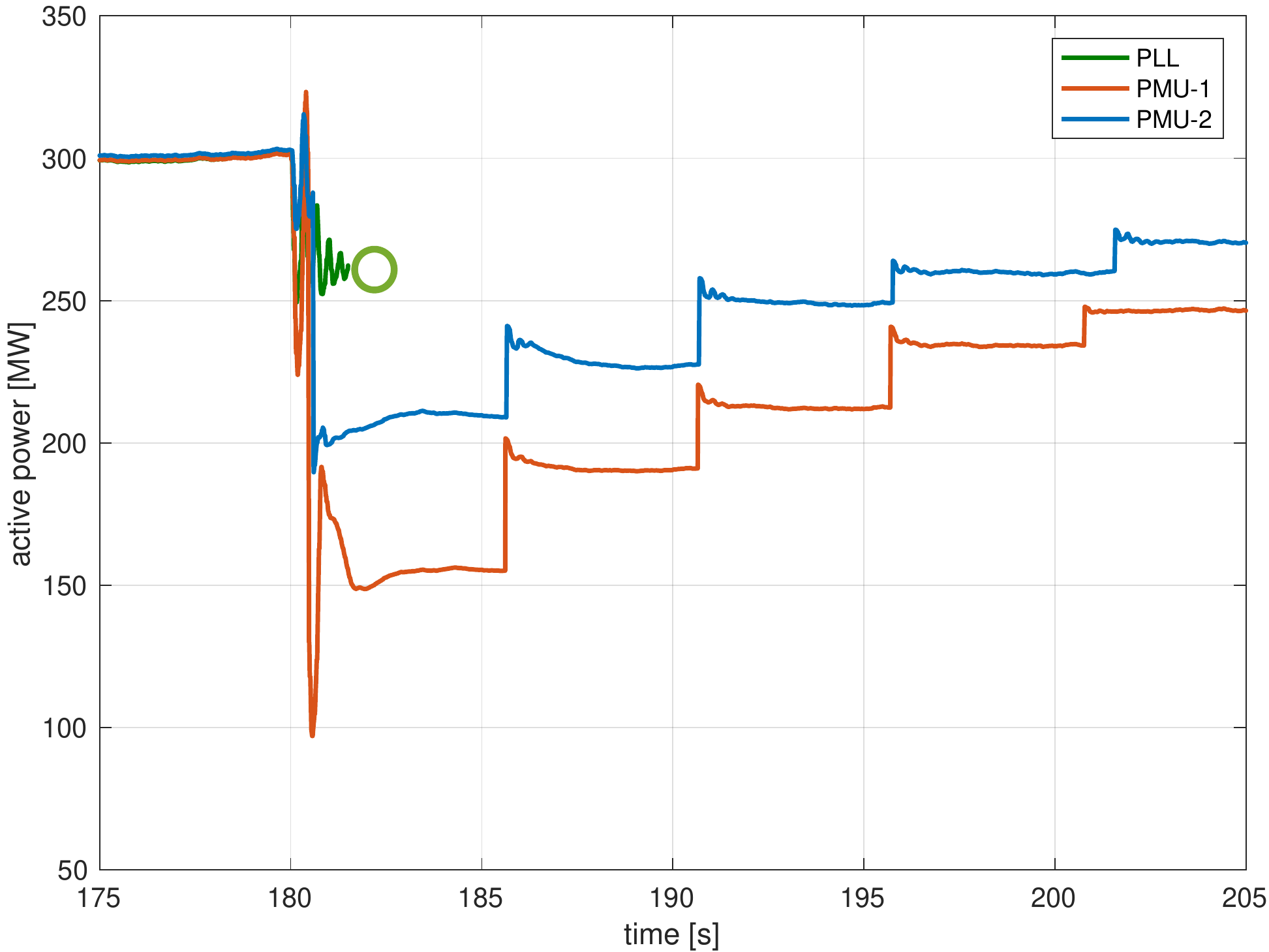}
		\caption{Power profile at load 23 as result of the UFLS based on PLL frequency measurements, as well as PMU-1 and PMU-2 ROCOF measurements in green, red and blue line, respectively. The green circle identifies the time instant when the entire system collapses due to the insufficient load-shedding action.}
		\label{fig:ufls_power}
	\end{figure}

	In Fig. \ref{fig:ufls_power}, instead, we show the active power profile at load 23 as result of the UFLS schemes based on PLL-based frequency and PMU-based ROCOF measurements. The frequency-based control scheme is not able to avoid the system blackout that occurs already at $ t $ = 182 s. Due to the closed-loop filtering stage, the PLL-based estimates of frequency prove to be excessively smoothed and delayed. As a consequence, the amount of shed loads is too limited, the system frequency rapidly decreases, and the protection relays trip the remaining generators.
	
	On the contrary, both ROCOF-based control schemes are able to restore the system stability, even if with different amounts of loads to be shed. In particular, PMU-2 allows for a more accurate ROCOF estimation and thus for an optimization of the load shedding scheme (e.g. avoiding unnecessary load disconnections). If we extend this analysis to the entire grid, we can characterize the algorithms' performance in terms of Expected Energy Not Served (EENS) that is equal to 13.87 and 11.56 MWh for PMU-1 and PMU-2, respectively.
	
	\section{Conclusions}
	In this paper, we investigate the feasibility of PMU-based ROCOF analysis, with specific reference to UFLS applications. To this end, we compare the estimation accuracy provided by four algorithms, relying on different signal models and ROCOF definitions. For this analysis, we consider two window lengths, namely 60 and 100 ms, representative of typical P- and M-class applications, respectively, and we evaluate their performance in numerically simulated scenarios, inspired by real-world acquisitions. As PMU-based measurements prove to be algorithm-dependent, a sensitivity analysis allows for identifying the most suitable modeling and estimation approach, that guarantees accurate and robust results even in dynamic operating conditions.
	
	The obtained results have experimentally validated that synchrophasor estimation algorithms guarantee enhanced accuracy, as long as the harmonic and inter-harmonic distortion within the measurement pass-band is reduced. Another critical condition is represented by transient events, when the synchrophasor model looses its appropriateness, and a PMU-based approach towards WAMPAC applications might become questionable. By contrast, PMUs provide optimal performance in the presence of slow fluctuations, as typical of inter-area oscillations.
	
	From this analysis it is possible to deduce some practical recommendations for PMU-based ROCOF measurements. These concepts can be directly inferred by the mathematical formulation of the adopted estimators, but have been rarely experimentally validated in a power system scenario. In particular, the employment of a dynamic signal model allows for a more accurate tracking of time-varying trends, like linear ramps or oscillations. On the other hand, a direct computation of the frequency first time-derivative proves to be gravely affected by spurious injections from measurement noise or uncompensated disturbances. Therefore, it is preferable to adopt a finite difference formulation that produces a smoothing effect on the final ROCOF estimates, at the cost of a slightly performance deterioration in the presence of transient events.
	
	In the last Section, we evaluate the actual feasibility of PMU-based ROCOF measurement by simulating a realistic scenario of UFLS application. As a term of comparison, we simulate also a more traditional control relying on the frequency estimate provided by a PLL. Based on the obtained results, the PMUs prove to be valuable alternatives to current nadir-based load-shedding relays as they allow for an accurate and prompt monitoring of the fundamental frequency and its first-time derivative. The predictive capability of ROCOF index allows for a more effective and prompt response to transient events, thus avoiding the occurrence of system blackouts and leading to a safe load restoration. The obtained results confirm that the adopted class (P or M) and signal model (static or dynamic) significantly affect the UFLS performance. Coherently with the previous analysis, the combination of M-class configuration and dynamic model provides the better results (i.e. the minimum EENS). 
	
	\section*{Acknowledgment}
	The Authors would like to thank Prof. Claudio Narduzzi (Universit\`a di Padova) and Prof. Alessandro Ferrero (Politecnico di Milano) for the interesting insights and discussions on ROCOF measurement validity and significance in the presence of non-sinusoidal conditions.

\end{document}